\documentclass[
  journal=pasa,
  manuscript=research-article, %article-type
  year=2025,
  volume=37
  ]{cup-journal}

\usepackage{amsmath}
\usepackage[nopatch]{microtype}
\usepackage{booktabs}

\usepackage{graphicx}	
\usepackage{xcolor}
\usepackage{array}
\usepackage{txfonts}

\usepackage{bm,booktabs}
\newcolumntype{N}{>{\centering\arraybackslash}m{.5in}}
\newcolumntype{G}{>{\centering\arraybackslash}m{\dimexpr2in+6\tabcolsep}}
\usepackage{relsize}
\usepackage{subcaption}
\usepackage{caption}

\usepackage{multirow}

\usepackage{hyperref}
\hypersetup{
    colorlinks = true,
    allcolors  = {blue} %Change to black if we want them invisible
}

\usepackage{placeins}

\newcommand{\Msun}{M$_{\odot}$} 
\newcommand{\Rsun}{R$_{\odot}$} 
\newcommand{\Lsun}{L$_{\odot}$}

\newcommand{\mesa}{\textsc{MESA}}
\newcommand{\mesastar}{\textsc{MESA~\!star}}

\newcommand{\Mdot}{M$_{\odot}\,\mathrm{yr}^{-1}$}

\newcommand{\FeH}{$[\rm{Fe}/\rm{H}]$}

\newcommand{\TiH}{$[\rm{Ti}/\rm{H}]$}

\newcommand{\referee}[1]{\textcolor{black}{#1}}

\title{Modelling the impact of circumbinary disk accretion on post-AGB binary evolution and surface chemistry}

\author{Kayla Martin}
\affiliation{School of Mathematical and Physical Sciences, Macquarie University, Balaclava Road, Sydney, NSW 2109, Australia}
\alsoaffiliation{Astrophysics and Space Technologies Research Centre, Macquarie University, Balaclava Road, Sydney, NSW 2109, Australia}
\email[Kayla Martin]{kayla.martin@hdr.mq.edu.au}

\author{Orsola De Marco}
\affiliation{School of Mathematical and Physical Sciences, Macquarie University, Balaclava Road, Sydney, NSW 2109, Australia}
\alsoaffiliation{Astrophysics and Space Technologies Research Centre, Macquarie University, Balaclava Road, Sydney, NSW 2109, Australia}

\author{Devika Kamath}
\affiliation{School of Mathematical and Physical Sciences, Macquarie University, Balaclava Road, Sydney, NSW 2109, Australia}
\alsoaffiliation{Astrophysics and Space Technologies Research Centre, Macquarie University, Balaclava Road, Sydney, NSW 2109, Australia}

\author{Glenn-Michael Oomen}
\affiliation{Institute of Astronomy, KU Leuven, Celestijnenlaan 200D bus 2401, B-3001, Leuven, Belgium}

\author{Hans Van Winckel}
\affiliation{Institute of Astronomy, KU Leuven, Celestijnenlaan 200D bus 2401, B-3001, Leuven, Belgium}

 %% First letter not capped
\begin{document}

\begin{abstract}
Post-asymptotic giant branch (post-AGB) binaries are surrounded by dusty circumbinary disks, and exhibit unexpected orbital properties resulting from poorly understood binary interaction processes. Re-accreted gas from the circumbinary disk alters the photospheric chemistry of the post-AGB star, producing a characteristic underabundance of refractory elements that correlates with condensation temperature \textendash\ a phenomenon known as chemical depletion. This work investigates how re-accretion from a disk drives chemical depletion, and the impact accreted matter has on post-AGB evolution. We used the \mesa\ code to evolve 0.55 and 0.60~\Msun\ post-AGB stars with the accretion of refractory element-depleted gas from a circumbinary disk. Our study adopts observationally-constrained initial accretion rates and disk masses to reproduce the chemical depletion patterns of six well-studied post-AGB binary stars: EP~Lyr, HP~Lyr, IRAS~17038-4815, IRAS~09144-4933, HD~131356, and SX~Cen. We find high accretion rates ($>10^{-7}$~\Mdot) and large disk masses ($\gtrsim10^{-2}$~\Msun) necessary to reproduce observed depletion, particularly in higher-mass, hotter post-AGB stars ($T_{\rm{eff}}\gtrsim$~6000~K). A slower evolution (lower core mass) is required to reproduce cooler ($T_{\rm{eff}}\lesssim$~5000~K) depleted post-AGB stars. Rapid accretion significantly impacts post-AGB evolution, stalling stars at cooler effective temperatures and extending post-AGB lifetimes by factors of around 3 to 10. Despite this, extended post-AGB timescales remain within or below the planetary nebula (PN) visibility timescale, suggesting accretion cannot account for the observed lack of ionised PNe in post-AGB binaries. Our findings constrain accretion-flow parameters and advance our understanding of disk-binary interactions in post-AGB systems. 
\end{abstract}

\section{Introduction}

%\referee{The asymptotic giant branch (AGB) is one of the final evolutionary stages for low- to intermediate-mass (0.8~-~8~\Msun) stars. Toward the end of the AGB phase, intense mass loss \textendash driven by stellar winds, binary interactions, or both \textendash rapidly reduces the envelope mass of the star down to $\lesssim$~0.02~\Msun\ \cite[]{Miller_Bertolami_2016}, on a timescale of $\sim10^{-6}-10^{-4}$~years \cite[]{Tosi_et_al_2022}. With such a low envelope mass, the outer stellar layers contract, causing the star to move off the AGB \cite[with a typical core mass of $\sim$~0.5~-~0.8~\Msun;][]{Miller_Bertolami_2016, Kamath_et_al_2023} and begin its post-AGB evolution. The post-AGB phase continues until the star heats to an effective temperature of $\sim$~25~000~K \cite[]{Schonberner_1987, VanWinckel_2003}, at which point it may ionise its circumstellar material to form a visible planetary nebula (PN). The duration of the post-AGB phase depends on the rate at which the remaining envelope mass is removed, with typical post-AGB timescales on the order of $10^3-10^4$~years \cite[]{Miller_Bertolami_2016}.}

\referee{The asymptotic giant branch (AGB) is one of the final evolutionary stages for low- to intermediate-mass (0.8~-~8~\Msun) stars. Toward the end of the AGB phase, stars with initial masses of $\sim$1~-~4~\Msun\ undergo intense mass loss \cite[$\sim10^{-6}-10^{-4}$~\Mdot;][]{Dellagli_et_al_2021, Tosi_et_al_2022} \textendash\ driven by stellar winds, binary interactions, or both \textendash\ which rapidly reduces their envelope mass to $\lesssim$~0.02~\Msun\ \cite[]{Miller_Bertolami_2016}. With such a low envelope mass, the outer stellar layers contract, causing the star to move off the AGB \cite[with a typical core mass of $\sim$~0.5~-~0.8~\Msun;][]{Miller_Bertolami_2016, Kamath_et_al_2023} and begin its post-AGB evolution. The post-AGB phase continues until the star heats to an effective temperature of $\sim$~25~000~K \cite[]{Schonberner_1987, VanWinckel_2003}, at which point it may ionise its circumstellar material to form a visible planetary nebula (PN). The duration of the post-AGB phase depends on the rate at which the remaining envelope mass is removed, with typical post-AGB evolution timescales on the order of $10^3-10^4$~years \cite[]{Miller_Bertolami_2016}.}

\referee{Many uncertainties still surround some of the most important physical processes undergone by stars on the dynamic AGB, including mass-loss, stellar winds, and pulsations. These uncertainties inevitably carry through to subsequent evolutionary stages, including the post-AGB, and are even further exacerbated when the star evolves in the presence of a binary companion; indeed, many post-AGB stars are found to evolve with a stellar companion, typically of mass $\sim$~1.1~$\pm$~0.6~\Msun\ \cite[]{Oomen_2018}. Consequently, the specific evolutionary pathway responsible for the unique characteristics observed in post-AGB binaries remains poorly defined.}

\referee{The spectral energy distributions (SEDs) of many post-AGB systems show a distinct near-infrared (near-IR) excess, indicative of a stable circumbinary disk composed of hot gas and dust in Keplerian rotation \cite[]{VanWinckel_2003, Kamath_2014, Kamath_2015, Kamath_et_al_2016, Kluska_2022}. The disks are highly extended, with radii $\sim$~100~-~500~AU \cite[]{Corporaal_2023b, corporaal_2023a, Andrych_et_al_2023, Katya_paper_2024}, and have a typical mass of $\sim 10^{-2}$~\Msun\ \cite[e.g.,][]{Bujarrabal_2013, Bujarrabal_et_al_2015, bujarrabal_2017, Bujarrabal_et_al_2018, Gallardo_et_al_2021}. The disks likely formed from material lost by the primary star during a phase of poorly understood binary interaction on the AGB, or possibly earlier, on the red-giant branch (RGB). The product of the latter are referred to as post-RGB binaries, distinguishable by their systematically lower luminosities \cite[]{Kamath_VanWinckel_2019}. Since post-AGB/RGB binaries are otherwise analogous, their unreliable luminosity estimates may lead to misclassifications. For this reason, such systems bearing circumbinary disks are more generally referred to as post-AGB binaries.}

%Around 85 Galactic post-AGB binaries have been identified to date \cite[see review by][]{VanWinckel_2019_review}, exhibiting a range of orbital eccentricities (up to $\sim$~0.65) and periods ($\sim$~100 - 3000~days). Such orbital characteristics are clear evidence of past interaction between the binary components, though the exact nature of the interaction mechanisms remains unclear \cite[e.g.,][]{Oomen_2018, Oomen_2020}. Further striking is the peculiar surface chemistry of the primary star, which displays a significant underabundance of refractory elements (e.g., Ti, Sc) that increases with condensation temperature \cite[$T_{\textrm{cond}}$;][]{lodders_2003}, while volatile elements (e.g., S, Zn) retain their main-sequence abundances. This phenomenon is referred to as `chemical depletion' \cite[see e.g.,][]{gezer_et_al_2015, Kamath_VanWinckel_2019, Oomen_2019, Maksym_2023a, Maksym_2023b}, and arises from the preferential re-accretion of \textit{clean} gas, resulting from the chemical fractionation of gas and dust in the disk \cite[]{Mosta_et_al_2019, Munoz_et_al_2019}, onto the post-AGB star, at typical rates of $10^{-7}-10^{-6}$~\Mdot\ \cite[]{Oomen_2019, Oomen_2020, bollen_2017,Bollen_2021,Bollen_2022}. Over time, the accreted refractory element-poor gas dilutes the photospheric composition of the star, resulting in metal-poor characteristics \cite[i.e., \FeH $\;=-5.0-0.5$~dex;][]{Rao_2014, gezer_et_al_2015, Kamath_VanWinckel_2019}.

Around 85 Galactic post-AGB binaries have been identified to date \cite[see review by][]{VanWinckel_2019_review}, exhibiting a range of orbital eccentricities (up to $\sim$~0.65) and periods ($\sim$~100 - 3000~days). Such orbital characteristics are clear evidence of past interaction between the binary components, though the exact nature of the interaction mechanisms remains unclear \cite[e.g.,][]{Oomen_2018, Oomen_2020}. Further striking is the significant underabundance of refractory elements (e.g., Ti, Sc) in the photosphere of the post-AGB primary star, which increases with condensation temperature \cite[$T_{\textrm{cond}}$;][]{lodders_2003}, while more volatile elements (e.g., S, Zn) retain their main sequence abundances. This phenomenon is referred to as `chemical depletion' \cite[see e.g.,][]{gezer_et_al_2015, Kamath_VanWinckel_2019, Oomen_2019, Maksym_2023a, Maksym_2023b}, and arises from the preferential re-accretion of \textit{clean} gas \textendash\ produced by the chemical fractionation of gas and dust \cite[]{Mosta_et_al_2019, Munoz_et_al_2019} \textendash\ from the disk onto the post-AGB star, at typical rates of $10^{-7}-10^{-6}$~\Mdot\ \cite[]{Oomen_2019, Oomen_2020, bollen_2017,Bollen_2021,Bollen_2022}. Over time, this refractory element-poor gas dilutes the star's photospheric composition, resulting in metal-poor characteristics \cite[i.e., \FeH $\;=-5.0-0.5$~dex;][]{Rao_2014, gezer_et_al_2015, Kamath_VanWinckel_2019}.

%Around 85 Galactic post-AGB binaries have been identified \cite[see review by][]{VanWinckel_2019_review}, exhibiting a range of orbital eccentricities (up to $\sim$0.65) and periods ($\sim$100 - 3000days), indicating past interaction between the binary components, though the interaction mechanisms remain unclear \cite[e.g.,][]{Oomen_2018, Oomen_2020}. 

%Notably, the primary star shows a significant underabundance of refractory elements (e.g., Ti, Sc) that increases with condensation temperature \cite[$T_{\textrm{cond}}$;][]{lodders_2003}, while volatile elements (e.g., S, Zn) retain their main-sequence abundances. 

%This "chemical depletion" \cite[see e.g.,][]{gezer_et_al_2015, Kamath_VanWinckel_2019, Oomen_2019, Maksym_2023a, Maksym_2023b} arises from the preferential re-accretion of \textit{clean} gas from the disk \cite[]{Mosta_et_al_2019, Munoz_et_al_2019}, at rates of $10^{-7}-10^{-6}$\Mdot\ \cite[]{Oomen_2019, Oomen_2020, bollen_2017,Bollen_2021,Bollen_2022}. Over time, this refractory element-poor gas dilutes the star's photospheric composition, leading to metal-poor characteristics \cite[i.e., \FeH $;=-5.0-0.5$~dex;][]{Rao_2014, gezer_et_al_2015, Kamath_VanWinckel_2019}.

\referee{Recently, \cite{Kluska_2022} categorised the circumbinary disks around Galactic post-AGB binaries into three main classes: i) full disks, which are rich in gas and dust, presenting a near-IR excess that extends outward from the dust sublimation radius \cite[]{hillen_2016, Kluska_et_al_2018, Kluska_2019}; ii) transition disks, which are observed in a fraction of post-AGB binaries (e.g., EP Lyr) and exhibit IR colours suggestive of a large (3~-~25~AU) dust-free inner-disk cavity \cite[]{corporaal_2023a}; and iii) dust-poor disks, which present a significant lack of IR excess, indicating very low or nearly absent circumbinary dust \cite[]{Maksym_2023b}. Post-AGB binaries with transition disks typically show stronger depletion of refractory elements \cite[]{Kluska_2022}.}

The observed chemical depletion patterns of post-AGB stars can be generally described by two key characteristics: i) the `depletion strength', measured as the abundance ratio of a non-depleted volatile element to a highly depleted refractory element, commonly $[\textrm{Zn}/\textrm{Ti}]$ \cite[]{gezer_et_al_2015}, and ii) the `turn-off temperature' ($T_{\textrm{turn-off}}$), corresponding to the condensation temperature beyond which elements are depleted. Observed depletion patterns are diverse \cite[]{Oomen_2019, Maksym_2023a, Maksym_2023b}, presenting a wide range of turn-off temperatures, from 800 to 1500~K, and varying depletion strengths, from mildly-to-moderately depleted if $[\textrm{Zn}/\textrm{Ti}]=0.5-1.5$~dex, to strongly depleted if $[\textrm{Zn}/\textrm{Ti}]>1.5$~dex. In the most extreme observed cases \cite[e.g., IRAS~11472-0800;][]{van_winckel_2012}, $[\textrm{Zn}/\textrm{Ti}]\!\approx$~4~dex, with a turn-off temperature of $\sim$~800~K. 

Observed depletion patterns display an underabundance of refractory elements that follows an approximately linear decline (in log space) up to high condensation temperatures (i.e., a `saturated' profile). Though in some cases, including several transition-disk objects (e.g., EP Lyr, DY~Ori, and GZ~Nor), refractory elements at higher condensation temperatures show similar underabundances (i.e., a `plateau' profile). Previous theoretical studies \cite[e.g.,][]{Oomen_2019,Oomen_2020}, based on abundances derived assuming local thermodynamic equilibrium (LTE), have successfully modelled observed depletion patterns, including the plateau profiles of transition-disk objects. However, non-LTE (NLTE) effects can significantly alter the abundance estimations in low-density regions, such as the outer layers of post-AGB stars, particularly of elements with resonance lines (e.g., Zn and Ti). As such, a recent study by \cite{Maksym_2023b} has presented NLTE-corrected abundances for a wide range of elements, and reveals a clear absence of plateau profiles in the transition disk subsample.

\referee{Importantly, gas accreted rapidly from a circumbinary disk extends the post-AGB lifetime of a star by replenishing its envelope mass, keeping the star cooler for longer and stalling its evolution until the disk mass is exhausted \cite[]{Miller_Bertolami_2016, Oomen_2019, Oomen_2020}. An extended post-AGB phase as a result of accretion can impact PN formation in post-AGB binaries, as circumstellar material may have sufficient time to expand before the star reaches ionisation temperature, resulting in a less strongly ionised or absent PN. Specifically, for PN formation to be possible, ionisation must occur within $\sim$~20~000~-~50~000~years \cite[the `PN visibility' timescale;][]{Schonberner_1987, PNE_exp_vel, PN_timescale, PN_visibility_2018}, depending on the stellar and ejected material properties. This imposes a limit on post-AGB evolution timescales, which must hence be shorter than the PN visibility timescale for the star to form a visible PN. Many evolved binaries predicted to be undergoing or to have recently undergone circumbinary disk re-accretion, typically lack an ionised PN. However, some exceptions exist, such as the emission nebulae around MyCin~18 \cite[]{MyCin18} and HD~101584 \cite[]{HD101584}, the reflection nebula around the Red Rectangle \cite[]{van_winckel_2014}, the nebula around IRAS~08005-2356 \cite[]{Manick_2021}, and the CO-bicone in 89~Her \cite[]{Bujarrabal_et_al_2007}.}

In this work, we use the Modules for Experiments in Stellar Astrophysics \cite[\mesa;][]{MESA_2011,MESA_2013,MESA_2015,MESA_2018,MESA_2019,MESA_update_2023} code to explore the impact of accreting chemically-depleted gas from a circumbinary disk onto evolving post-AGB stars. Specifically, we model the chemical depletion process as in \cite{Oomen_2019,Oomen_2020}, to reproduce the observed depletion patterns of six post-AGB binary stars, namely EP~Lyr, HP~Lyr, IRAS~17038-4815, IRAS~09144-4933, HD~131356, and SX~Cen. Further, we assess how accretion influences the structural and evolutionary properties of post-AGB binary stars, and their ability to form ionised PNe. In turn, we aim to refine our understanding of the accretion-flow parameters for disk-binary interactions during the post-AGB phase, to aid future studies.

\section{Data}
\label{postAGB_sample}

We selected a sample of six disk-type post-AGB binary stars, with well-studied orbital parameters \cite[see e.g.,][]{Oomen_2018} and readily available chemical abundance data: EP~Lyr, HP~Lyr, IRAS~17038-4815, IRAS~09144-4933, HD~131356, and SX~Cen. The important stellar and binary properties for each star are summarised in Table~\ref{star_prop_table}. The SEDs for the target sample are further provided in \ref{Appendix_SEDs}.

\begin{table*}[hbt!]
    \centering
    \caption{Important stellar and binary properties for our post-AGB sample. \referee{Included are the post-AGB star's effective temperature ($T_{\rm{eff}}$) and SED-derived photospheric luminosity \protect\cite[from][]{Oomen_2019}}, as well as estimated current mass and approximate stellar radius. Orbital periods, eccentricities, and projected semi-major axes ($a_{\rm{b}}\sin i$), are from \protect\cite{Oomen_2018,Oomen_2020}. The upper-limit Roche-lobe radii at periastron ($R_{\rm{L,\,peri}}$) is additionally given, computed using minimum companion masses from \protect\cite{Oomen_2018}. Further provided are key chemical abundance ratios (in dex, relative to solar), depletion pattern turn-off temperature ($T_{\rm{turn-off}}$), and circumbinary disk type \protect\cite[from][]{Kluska_2022}. Reference papers are provided below the table.}
    \setlength{\tabcolsep}{5pt}
    \renewcommand{\arraystretch}{0.8}
    \resizebox{\textwidth}{!}{\begin{tabular}{p{28mm} >{\centering\arraybackslash}p{20mm} >{\centering\arraybackslash}p{20mm} >{\centering\arraybackslash}p{24mm} >{\centering\arraybackslash}p{24mm} >{\centering\arraybackslash}p{22mm} >{\centering\arraybackslash}p{20mm}}
         \hline
         \hline 
         \rule{0pt}{1mm}Star Name & EP Lyr & HP Lyr & IRAS~17038-4815 & IRAS~09144-4933 & HD~131356 & SX Cen \\
         \hline 
         
         $T_{\rm{eff}}$ (K)  & $6200$ & $6300$ & $4750$ & $5750$ & $6000$ & $6250$ \\
         Luminosity ($10^3$~\Lsun)  & $5.5\,^{+2.3}_{-1.4}$ & $4.9\,^{+2.0}_{-1.3}$ & $4.8\,^{+3.2}_{-1.6}$ & $4.2\,^{+1.7}_{-1.1}$ & $3.5\,^{+1.0}_{-0.7}$ & $3.3\,^{+1.3}_{-0.9}$ \\
         %Current mass (\Msun)  & $0.60\,^{+0.04}_{-0.02}$ & $0.59\,^{+0.04}_{-0.02}$ & $0.58\,^{+0.06}_{-0.03}$ & $0.57\,^{+0.03}_{-0.02}$ & $0.56\,^{+0.02}_{-0.01}$ & $0.56\,^{+0.02}_{-0.02}$ \\
         \referee{Current mass (\Msun)}  & \referee{$0.60$} & \referee{$0.59$} & \referee{$0.58$} & \referee{$0.57$} & \referee{$0.56$} & \referee{$0.56$} \\      
         Radius (\Rsun)  & $65\pm13$ & $59\pm11$ & $103\pm33$ & $66\pm12$ & $55\pm6$ & $49\pm9$ \\
         Period (days)              & $1151\pm14$   & $1818\pm80$   & $1394\pm12$   & $1762\pm27$   & $1488.0\pm8.7$ & $564.3\pm7.6$ \\
         Eccentricity               & $0.39\pm0.09$ & $0.20\pm0.04$ & $0.63\pm0.06$ & $0.30\pm0.04$ & $0.32\pm0.04$  & $0.00\pm0.06$ \\
         $a_{\rm{b}}\sin i$ (AU) & $1.30\pm0.12$ & $1.27\pm0.06$ & $1.52\pm0.08$ & $2.25\pm0.11$ & $2.11\pm0.09$ & $1.12\pm0.05$ \\  
         $R_{\rm{L,\,peri}}$ (\Rsun)  & $75\pm15$ & $79\pm11$ & $49\pm12$ & $152\pm19$ & $140\pm16$ & $110\pm13$  \\
         
         $[\rm{Fe}/\rm{H}]$       & $-1.8$ & $-1.0$ & $-1.5$ & $-0.3$ & $-0.7$ & $-1.2$ \\      
         $[\rm{Zn}/\rm{H}]$       & $-0.7$ & $-0.4$ & $-1.3$ & /      & $-0.6$ & $-0.5$ \\
         $[\rm{Ti}/\rm{H}]$       & $-2.0$ & $-3.0$ & $-2.0$ & $-1.3$ & $-1.1$ & $-2.0$ \\
         $[\rm{S}/\rm{H}]$        & $-0.6$ & $0.1$  & /      & $0.0$  & $-0.6$ & $-0.1$ \\
         $[\rm{Zn}/\rm{Ti}]$      & $1.3$  & $2.6$  & $0.7$  & /      & $0.6$  & $1.4$  \\
         $[\rm{S}/\rm{Ti}]$       & $1.4$  & $3.0$  & /      & $1.3$  & $0.5$  & $1.9$  \\
         %Depletion strength               & Moderate & Strong & Mild & Moderate & Mild & Strong \\
         $T_{\rm{turn-off}}$ (K)      & $800$ & $1300$ & $1400$ & $1400$ & $1300$ & $1100$ \\
         Disk type                         & \rm{Transition} & \rm{Full} & \rm{Full} & \rm{Full} & \rm{Full} & \rm{Full} \\   

         Reference               & $1,2,3$ & $2,4$ & $4,5$ & $4,5$ & $6$ & $7$ \\%[1mm]
         \hline
         
         \multicolumn{7}{p{0.99\linewidth}}{\textbf{Notes.} Formal errors on temperature are $\pm250$~K, and for abundances $\pm0.3$~dex. Depletion pattern turn-off temperatures are approximate.}\\
         \multicolumn{7}{p{0.99\linewidth}}{\textbf{References.} $(1)$ \cite{Gonzalez_1997}; $(2)$ \cite{manick_2017}; $(3)$ \cite{Maksym_2023b}; $(4)$ \cite{Giridhar_2005}; $(5)$ \cite{Maas_et_al_2005}; $(6)$ \cite{Van_Winckel_1997}; $(7)$ \cite{Maas_et_al_2002}.} \\
    \end{tabular}}
    \label{star_prop_table}
\end{table*}

\referee{The effective temperature and photospheric luminosity for each post-AGB star, retrieved from \cite{Oomen_2019,Oomen_2020}, are provided in Table~\ref{star_prop_table}. Luminosities were derived through SED-fitting using \textit{Gaia} DR2 distances; see \cite{Oomen_2018, Oomen_2019} for details on the SED fitting procedure.} Since these distances rely on single-star astrometric solutions, the corresponding luminosities, and hence, current mass estimates, are unreliable, particularly for RV Tauri pulsators EP~Lyr, HP~Lyr, IRAS~17038-4815, and IRAS~09144-4933. We note that SED-fits based on \textit{Gaia} DR3 distances \cite[see][]{GaiaDR3_dist_BailerJones2021} were not readily available for the post-AGB sample at the time of this study; for completeness, an example calculation using \textit{Gaia} DR3 is provided in \ref{Appendix_GDR3} for EP Lyr and HP Lyr. \referee{The current mass of each post-AGB star in Table~\ref{star_prop_table} was estimated using the core mass-luminosity relation of \cite{stellar_wind}, and the current radius from the star's luminosity and effective temperature.} The upper-limit Roche-lobe radius at periastron in Table~\ref{star_prop_table} was computed with the standard \cite{eggleton_1983} formula, assuming estimated companion masses from \cite{Oomen_2018,Oomen_2020}.

\section{Modelling post-AGB binary evolution}

The post-AGB models presented in this work were computed using version \texttt{r24.03.1} of the open-source stellar structure and evolution code \mesa\ \cite[]{MESA_2011,MESA_2013,MESA_2015,MESA_2018,MESA_2019,MESA_update_2023}. \mesa's one-dimensional stellar evolution module, \mesastar, builds a spherically symmetric model from a mesh of several thousand cells, configured as concentric shells. \mesastar~evolves the stellar model by solving a set of fully-coupled structure and composition equations at each point along the stellar radius, updating the mesh structure and composition profiles at each timestep. 

To simulate post-AGB evolution with the accretion of refractory element-poor gas from a circumbinary disk, we adopted the approach of \cite{Oomen_2019,Oomen_2020}. To remain concise, we describe here only the most important model considerations for our study, and refer the reader directly to the previous works for all other model details and specifications.

\subsection{Simulating accretion onto a post-AGB binary star}
\label{postAGBmodelling_section}

After some strong interaction with a companion, the envelope of the primary AGB star is rapidly removed, causing the star to contract and begin its post-AGB evolution toward higher temperatures. The post-AGB star then detaches from its Roche lobe (RL), after which point we do not expect further interaction between the stars. Post-AGB evolution may therefore be reasonably approximated by single star \mesa\ models. 

%To create the post-AGB models for this work, we evolved a star of initial mass $2.5$~\Msun, and solar-scaled metallicty \cite[i.e., $Z_0 =$~0.02, with composition from][]{asplund_2009}, from the main-sequence up the AGB until a He-core mass of 0.55 and 0.60~\Msun\ \cite[i.e., the typical masses of PN-forming post-AGB stars;][]{Miller_Bertolami_2016}. At this point, we simulated a strong interaction with a companion star by artificially increasing the stellar mass loss to $10^{-4}$~\Mdot\ \cite[as in][]{Oomen_2019}, rapidly removing the star's envelope down to $\sim$~0.02~\Msun\ on a short timescale of 10~000~years, pushing the star onto the post-AGB. We note that approximately 20\% of stars departing the AGB during a thermal pulse cycle experience a late thermal pulse (LTP), which reignites helium burning and impacts the stellar evolution. In the 0.55~\Msun\ models created for this study, such an LTP occurs around 8000~years into the post-AGB phase, and is explored further in \ref{Appendix_LTP}.

To create the post-AGB models for this work, we evolved a star of initial mass $2.5$~\Msun, and solar-scaled metallicty \cite[$Z_0 =$~0.02, with composition from][]{asplund_2009}, from the main-sequence through the AGB phase, until a He-core mass of 0.55 and 0.60~\Msun. These core masses represent the approximate lower and upper bounds of our chosen post-AGB star sample (see Table 1) and align with the typical masses of PN-forming post-AGB stars \cite[]{Miller_Bertolami_2016}. At this point, we simulated a strong interaction with a companion by artificially increasing the mass-loss rate to $10^{-4}$~\Mdot\ \cite[as in][]{Oomen_2019}, rapidly removing the envelope down to $\sim$~0.02~\Msun\ over 10~000~years, pushing the star onto the post-AGB. 

\referee{Such rapid mass loss, from the already thin AGB envelope, inhibits thermal pulses and prevents carbon enrichment via the third dredge-up \cite[see e.g.,][]{Kamath_et_al_2023}, resulting in O-rich surface chemistry (C/O $\sim$ 0.29) in both the 0.55~\Msun\ and 0.60~\Msun\ post-AGB models. Since we do not consider surface abundance changes in C, N, or O in our depletion analysis (see Section~\ref{depl_method}), whether our stars are C- or O-rich when leaving the AGB is not crucial.} Additionally, around 20\% of stars departing the AGB during a thermal pulse cycle experience a late thermal pulse (LTP), which reignites helium burning and impacts stellar evolution \referee{\cite[see][]{Schonberner_1979, Iben_et_al_1983, Blocker_2001, Herwig_2001}}. An LTP occurs $\sim$~8000~years into the post-AGB phase of the 0.55~\Msun\ models created for this study, and is explored further in \ref{Appendix_LTP}.

Once on the post-AGB, we evolved the star with the accretion of gas, presumed to originate from a circumbinary disk. Accretion was applied from $T_{\rm{eff}}=$~4000~K, which corresponds to the approximate lower bound on the predicted temperature range for $\sim$~0.55~-~0.60~\Msun\ stars entering the post-AGB phase \cite[]{Miller_Bertolami_2016, Kamath_et_al_2023}. We note that due to the large radii of stars entering the post-AGB, rapid accretion may temporarily expand the star's outer layers. If the star exceeds its RL radius, mass-transfer may ensue, having potential consequences on e.g., envelope mass and surface chemistry. However, since most of our post-AGB systems have wide orbital separations ($>$~1000~days), and hence large RL radii (see Table~\ref{star_prop_table}), such additional mass-transfer would have minimal impact on our results. 

To model accretion we adopted the time-dependent mass-accretion rate from \cite{Oomen_2019}, expressed as
\begin{equation}
    \label{acc_eq}
    \dot{M}_{\rm{acc}}(t) = \dot{M}(0) \left( 1 + \dfrac{4\dot{M}(0)t}{M_{\rm{d}}} \right)^{-3/2} \; .
\end{equation}
Equation~\ref{acc_eq} describes the accretion flow from a viscously-evolving circumbinary disk using two separate parameters: the initial accretion rate, $\mbox{$\dot{M}(t\!=\!0)$}$, and initial disk mass, $M_{\rm{d}}$. Here, initial accretion rates of $10^{-8}$, $10^{-7}$, $5\times10^{-7}$, and $10^{-6}$~\Mdot\ were used, to encompass the typical mass accretion rate derived for post-AGB binaries \cite[i.e., $\sim\!10^{-7}$~\Mdot;][]{Izzard_Jermyn_2018, Oomen_2019}. Similarly, initial disk masses of $7\times10^{-3}$, $10^{-2}$, and $3\times10^{-2}$~\Msun, were selected to align with the most typical observed post-AGB disk masses \cite[i.e., $\sim\!10^{-2}$~\Msun;][]{Bujarrabal_2013, Hillen_2017, Kluska_et_al_2018}.  

The expression given in Equation~\ref{acc_eq} assumes the post-AGB star accretes approximately half of the gas entering the binary cavity, while the rest is captured by the companion. The rate of accretion onto the post-AGB star is hence half the mass-loss rate from the disk, leading to an initial viscous evolution time of $t_0 = M_{\rm{d}}\,/\,4\dot{M}(0)$ for the disk. The accretion rate of Equation~\ref{acc_eq} decays non-linearly from its initial value, scaling with the disk density which decreases slowly over time due to viscous expansion as a result of mass loss \cite[]{Rafikov_2016}.

\subsection{Modelling the chemical depletion process}
\label{depl_method}

\referee{Chemical depletion is observed in most post-AGB binaries with circumbinary disks, and is attributed to interactions, namely accretion, between the disk and post-AGB primary star. Due to the low envelope mass of post-AGB stars ($\lesssim$~0.02~\Msun), the relative abundances, $[\rm{X}/\rm{H}]$\footnote{\referee{$[\rm{X}/\rm{H}]=\log \dfrac{N(X)}{N(H)}-\log \dfrac{N(X)_{\odot}}{N(H)_{\odot}}$}, where $N(X)$ and $N(H)$ are the number abundances of element $X$ and hydrogen, respectively. The subscript $\odot$ represents the corresponding solar value.}, of elements in the stellar photosphere are easily dominated by the refractory element-poor gas accreted from the disk. This results in a unique photospheric depletion pattern, characterised by an underabundance of refractory elements (e.g., Ti, Sc), while more volatile elements (e.g., S, Zn) retain their main-sequence values.}

\referee{To investigate these depletion patterns, we model the chemical evolution of six post-AGB binary stars: EP~Lyr, HP~Lyr, IRAS~17038-4815, IRAS~09144-4933, HD~131356, and SX~Cen. The predicted current masses (Table~\ref{star_prop_table}) of the target sample range from $\sim$~0.55 to 0.60~\Msun, making them viable candidates for PN formation. Further, these stars have readily available chemical abundance data, and exhibit depletion patterns of varying strengths ($[\rm{Zn}/\rm{Ti}]=$~0.6~-~2.6~dex) and turn-off temperatures ($\sim$~800~-~1400~K). The chemical abundance data for each post-AGB star has been sourced from the relevant works listed in Table~\ref{star_prop_table}. For EP~Lyr, we employ NLTE-corrected abundances for elements C, N, O, S, Na, Mg, Al, Si, Ca, and Fe from \cite{Maksym_2023b}, though note that such corrections are not yet available for the remaining stars in the sample.}
 
%To trace chemical depletion, we use $[\rm{Zn}/\rm{Ti}]$, derived from the relative abundances $[\rm{Zn}/\rm{H}]$ and $[\rm{Ti}/\rm{H}]$ in the outer layers of our models. Where no observed Zn abundance is available (i.e., IRAS~09144-4933), we use $[\rm{S}/\rm{Ti}]$ to trace chemical depletion \cite[as in][]{Oomen_2019}. Our depletion modelling framework was based on the most depleted observed post-AGB stars, with \TiH$\,\approx-4$~dex \cite[e.g., IRAS~11472-0800;][]{van_winckel_2012}, such that any star with $[\rm{Zn}/\rm{Ti}]<$~4~dex may be reasonably reproduced by our models. Specifically, the initial composition of the accretion-flow was set such that the abundances of elements with condensation temperatures higher than Zn (726~K), and up to Ti (1582~K), were linearly interpolated \cite[as in][]{Oomen_2019} down to a maximum depletion strength of $[\rm{Zn}/\rm{Ti}]=$~4~dex. The highly volatile elements C, N, O, and S, were normalised to their abundances upon entering the post-AGB phase, since any deviations from their main-sequence values are due to RGB and AGB nucleosynthetic and mixing processes not crucial to this study \cite[see][]{Ventura_2008, Ventura_2018, Ventura_2020, Kobayashi_2020, Kamath_et_al_2023, Maksym_2023a, Meghna_2024}.

\referee{We trace chemical depletion as $[\rm{Zn}/\rm{Ti}]$, using the relative abundances $[\rm{Zn}/\rm{H}]$ and $[\rm{Ti}/\rm{H}]$ in the outer layers of the models. Where no observed Zn abundance was available (i.e., IRAS~09144-4933), $[\rm{S}/\rm{Ti}]$ was used to trace depletion, though we note that weaker spectral lines and larger uncertainties in abundance measurements make S a less reliable depletion tracer than elements like Zn and Ti in post-AGB stars. Further, we exclude C, N, and O from our depletion analysis as their abundances are primarily governed by RGB and AGB nucleosynthesis, consistent with single-star evolutionary processes, and remain largely unaffected by binary-related interactions \cite[see][]{Ventura_2008, Ventura_2018, Ventura_2020, Kobayashi_2020, Kamath_et_al_2023, Maksym_2023a, Meghna_2024}.}

Our depletion modelling framework was based on the most depleted observed post-AGB stars, with \TiH$\,\approx-4$~dex \cite[e.g., IRAS~11472-0800;][]{van_winckel_2012}, such that any star with $[\rm{Zn}/\rm{Ti}]<$~4~dex may be reproduced by our models. Specifically, the initial composition of the accretion-flow was set such that the abundances of elements with condensation temperatures higher than Zn (726~K), and up to Ti (1582~K), were linearly interpolated \cite[as in][]{Oomen_2019} down to a maximum depletion strength of $[\rm{Zn}/\rm{Ti}]=$~4~dex. \referee{The predicted depletion pattern was then scaled to the observed abundance of Zn (or S for IRAS~09144-4933), assumed to reflect the initial metallicity of the star \cite[]{Oomen_2019}.}

To ensure a smooth evolution of depletion and prevent accreted gas from accumulating in the outer cells of the model, the accreted gas was mixed through the star's convective outer layers down to regions where $T\lesssim$~80~000~K (for more details, see \ref{Appendix_Menv}). We note that while accreted material may mix to varying depths in post-AGB stars, \cite{Oomen_2019} showed that this has minimal impact on depletion, particularly at lower effective temperatures ($\lesssim$~7000~K), for which the outer layers of post-AGB stars are still convective. Convective mixing was treated using the Mixing Length Theory \cite[]{Joyce_and_Tayar_2023}, with a semiconvective mixing efficiency of $\alpha=$~0.04, and a thermohaline mixing coefficient of 2.0. As in \cite{Ventura_2018} and \cite{Kamath_et_al_2023}, convective overshoot was treated using an exponential scheme over an e-folding distance of 0.02 times the scale height. For comprehensive details of the depletion modelling, see \cite{Oomen_2019}.

\subsection{Stellar winds}

An important physical consideration in our models was the treatment of stellar winds, which directly influence a star's envelope mass and hence post-AGB evolution. As in \cite{Oomen_2019}, wind-driven mass loss on the RGB and early AGB ($T_{\rm{eff}}\lesssim$~5000~K) was described using the semi-empirical, \cite{Reimers}-type wind prescription from \cite{Updated_Reimers}, referred to as the \textsc{SCwind}, which is expressed as
\begin{equation}
    \label{SCwind_eq}
    \dot{M}_{\rm{SCwind}} = \; \eta \; \cdot \, \dfrac{L_{*}R_{*}}{M_{*}} \; \left( \dfrac{T_{\rm{eff}}}{4000\rm{K}}  \right) ^{3.5} \left( 1+ \dfrac{g_{\odot}}{4300 g_{*}} \right) \; ,
\end{equation}
where $L_{*}$, $R_{*}$, $M_{*}$, and $T_{\rm{eff}}$ are the star's luminosity, radius, mass, and effective temperature, respectively, $g_{\odot}$ is solar surface gravity, and scaling parameter $\eta=8 \times 10^{-14}$~\Mdot. The SCwind describes cool stellar winds, and yields typical mass-loss rates of $\sim\!10^{-6}$~\Mdot\ \cite[]{Updated_Reimers}. Typically, Equation~\ref{SCwind_eq} is calibrated for RGB stars, where stellar winds are driven by chromospheric activity and strong magnetic fields absent in AGB and post-AGB stars \cite[]{Sabin_2015}. However, we currently lack a better alternative for post-AGB modelling at lower temperatures. Strong pulsation-enhanced, dust-driven winds during the AGB thermal pulse stage were further accounted for by scaling the SCwind by the stellar pulsation period.

To describe radiation-driven stellar winds at high effective temperatures ($\gtrsim$~8000~K) on the post-AGB, the \cite{Blocker}-type wind prescription from \cite{Miller_Bertolami_2016} was adopted. This prescription is typically calibrated for hot ($T_{\rm{eff}}\gtrsim$~30~000~K) central stars of PNe (CSPNe), and is hence referred to as the \textsc{CSPNwind}, expressed as 
\begin{equation}
    \label{CSPNe_wind_eq}
    \dot{M}_{\rm{CSPNwind}} = \; \zeta \; \cdot \; \left( \dfrac{L_{*}}{L_{\odot}}  \right) ^{1.674} \left( \dfrac{Z_{0}}{Z_{\odot}} \right)^{2/3} \; ,
\end{equation}
where $Z_0$ is the star's initial metallicity, and scaling parameter $\zeta=9.778 \times 10^{-15}$~\Mdot. The CSPNwind yields typical mass-loss rates of $\sim\!10^{-9}$~\Mdot. As this is much longer than the nuclear burning rate of hydrogen at the convective envelope base ($\sim\!10^{-7}-10^{-5}$~\Mdot, depending on luminosity), mass-loss by the CSPNwind (Equation~\ref{CSPNe_wind_eq}) has negligible impact on the evolution of hotter post-AGB stars \cite[]{Oomen_2019}. 

Following the approach of \cite{Oomen_2019}, we defined a transition region over which mass-loss rates exponentially decrease from those of the cool SCwind (Equation~\ref{SCwind_eq}) to those of the hot, CSPNwind (Equation~\ref{CSPNe_wind_eq}). Specifically, we define this transition to occur over the effective temperature range of 5000~-~8000~K, during which the SCwind is predicted to subside \cite[]{Cranmer_2011, Miller_Bertolami_2016}.

\section{Results: chemical depletion}
% ---- NOTES ---- 
% - [Zn/Ti] determines how depleted a star is: 0.5-1.0 is mild, 1.0-1.5 is moderate, >1,5 is strong
% - [Fe/Ti] indicates whether or not the depletion profile is saturated: saturated if >0.25, plateaued if <0.25
% --------------

In order to reproduce the observed chemical depletion of post-AGB stars at specific effective temperatures, the depletion process must occur on a similar, or faster timescale than the post-AGB evolution of the star. The depletion timescale ($\mathlarger{\mathlarger{\tau}}_{\rm{depl}}$) describes how rapidly the photospheric abundances of a star will become dominated by the accreted disk gas abundances, and is largely determined by the initial accretion rate, $\dot{M}(0)$, and disk mass, $M_{\rm{d}}$, as $\mathlarger{\mathlarger{\tau}}_{\rm{acc}}\propto M_{\rm{d}}/\dot{M}(0)$.

In this section, we present the \mesa\ models used to explore the formation of chemical depletion patterns in post-AGB binary primary stars with current mass $\sim$~0.55~-~0.60~\Msun. Further, we compare the chemical depletion patterns predicted by our models with those observed in six post-AGB binary stars, namely EP~Lyr, HP~Lyr, IRAS~17038-4815, IRAS~09144-4933, HD~131356, and SX~Cen, which exhibit diverse depletion pattern properties (see Table~\ref{star_prop_table}). In turn, we provide the suitable range of accretion rates and circumbinary disk masses to reproduce chemical depletion in post-AGB binary stars.

\subsection{Refining the required accretion parameters to reproduce observed chemical depletion}

\begin{figure}[hbt!]
    \centering
    \includegraphics[width=0.98\linewidth]{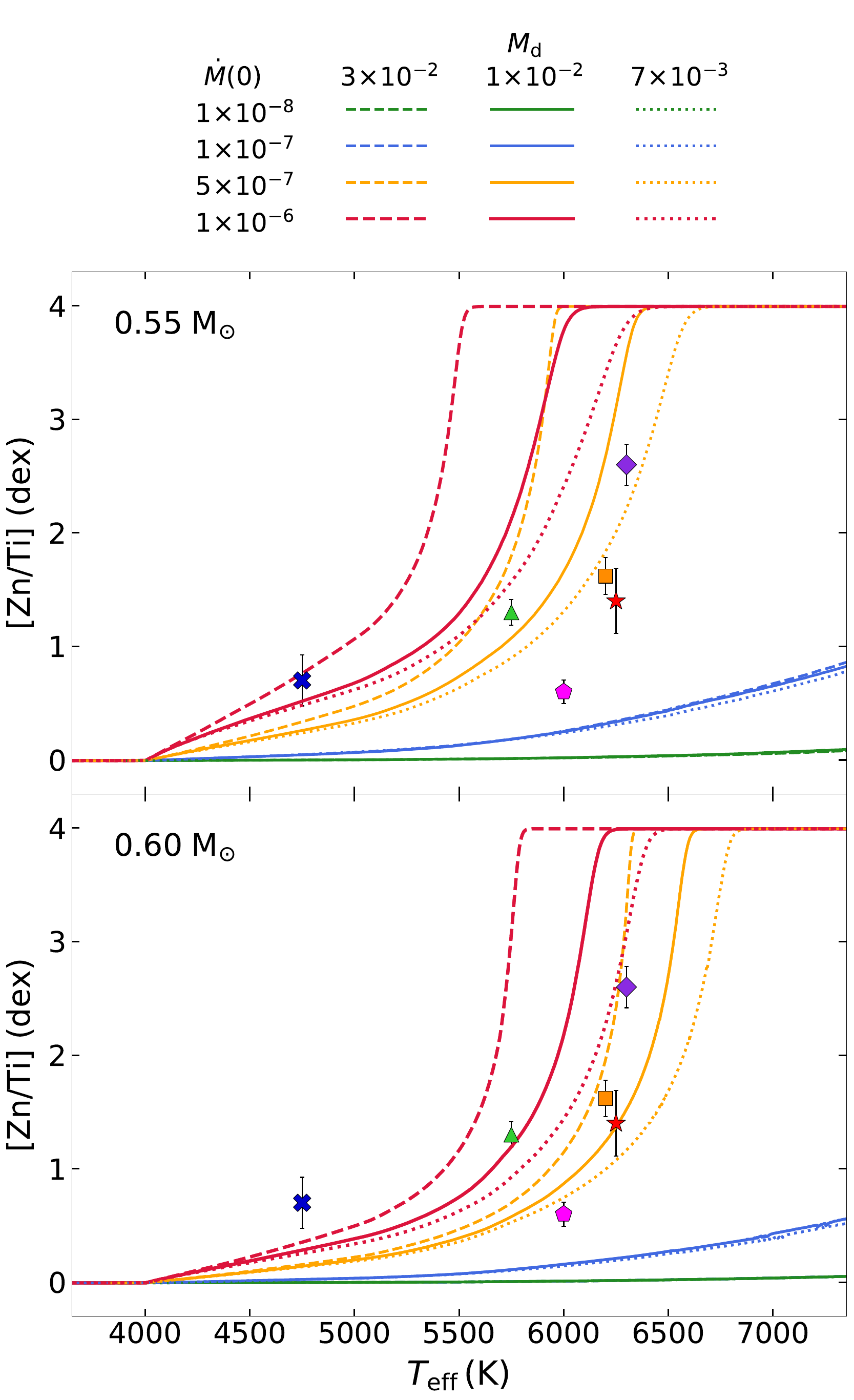}
    \caption{Post-AGB \mesa\ models of mass 0.55~\Msun\ (top) and 0.60~\Msun\ (bottom), with maximum depletion strength $[\rm{Zn}/\rm{Ti}]=$~4~dex. Models begin at 4000~K, and increase in effective temperature over time. Different initial accretion rates, $\dot{M}(0)$ (in \Mdot), are represented by different coloured curves, while different linestyles corresponds to different disk masses, $M_{\rm{d}}$ (in \Msun). Observed stars are over-plotted as points: EP~Lyr (yellow square), HP~Lyr (purple diamond), IRAS~17038-4815 (blue cross), IRAS~09144-4933 (green triangle), HD~131356 (pink pentagon), and SX~Cen (red star).}
    \label{ZnTi_Teff}
\end{figure}

To reproduce the depletion patterns of specific post-AGB stars, the observed depletion strength ($\left[\rm{Zn}/\rm{Ti}\right]$) must be achieved in the model by the star's effective temperature (see Table~\ref{star_prop_table}). The initial accretion rates and disk masses capable of reproducing depletion in a given post-AGB star may hence be determined by its position in $\left[\rm{Zn}/\rm{Ti}\right]-T_{\rm{eff}}$ space \cite[]{Oomen_2019}. This is demonstrated in Figure~\ref{ZnTi_Teff} for the six post-AGB stars in our sample (see Section~\ref{postAGB_sample}). Since the disk model used in this work is based on the most depleted circumbinary disks (with $\left[\rm{Zn}/\rm{Ti}\right]=$~4~dex), depletion in any post-AGB star falling below a particular curve in $\left[\rm{Zn}/\rm{Ti}\right]-T_{\rm{eff}}$ space may be reproduced by that combination of accretion parameters. The best-fitting models from Figure~\ref{ZnTi_Teff} hence provide only lower limits on the accretion-flow conditions required to reproduce depletion in post-AGB stars: the best-fitting accretion parameters from Figure~\ref{ZnTi_Teff} are summarised in Table~\ref{bestfit_acc_params_table} for the target sample. 

\begin{table}[hbt!]
    \centering
    \caption{Best-fitting initial accretion rates, $\dot{M}(0)$ (in \Mdot), and disk masses, $M_{\rm{d}}$ (in \Msun), per core mass ($M_{\rm{c}}$), determined from the position of each post-AGB star in $[\rm{Zn}/\rm{Ti}]-T_{\rm{eff}}$ space (see Figure~\ref{ZnTi_Teff}).}
    \setlength{\tabcolsep}{3.2pt}
    \renewcommand{\arraystretch}{1.0}
    \resizebox{\linewidth}{!}{\begin{tabular}{p{23mm} >{\centering\arraybackslash}p{13mm} >{\centering\arraybackslash}p{13mm} p{1mm} >{\centering\arraybackslash}p{13mm} >{\centering\arraybackslash}p{13mm}}
         \hline
         \hline 
          & \multicolumn{2}{c}{$M_{\rm{c}}=0.55$~\Msun} & & \multicolumn{2}{c}{$M_{\rm{c}}=0.60$~\Msun} \\
         \cline{2-3} \cline{5-6}
         \rule{0pt}{1mm}Star Name & $\dot{M}(0)$ & $M_{\rm{d}}$ & & $\dot{M}(0)$  & $M_{\rm{d}}$\\
         \hline  
         EP Lyr        & $5\times10^{-7}$  & $7\times10^{-3}$  & & $5\times10^{-7}$  &  $3\times10^{-2}$ \\
         HP Lyr        & $5\times10^{-7}$  & $1\times10^{-2}$  & & $5\times10^{-7}$  &  $3\times10^{-2}$ \\
         IRAS~$17038\rm{-}4815$  & $1\times10^{-6}$  & $3\times10^{-2}$  & & $1\times10^{-6}$  &  $3\times10^{-2}$ \\
         IRAS~$09144\rm{-}4933$  & $1\times10^{-6}$  & $7\times10^{-3}$  & & $1\times10^{-6}$  &  $3\times10^{-2}$ \\
         HD~$131356$   & $5\times10^{-7}$  & $7\times10^{-3}$  & & $5\times10^{-7}$  &  $7\times10^{-3}$ \\
         SX Cen        & $5\times10^{-7}$  & $7\times10^{-3}$  & & $5\times10^{-7}$  &  $1\times10^{-2}$ \\
         \hline
         \multicolumn{6}{p{\linewidth}}{\textbf{Note.} Corresponds to accretion timescales, $M_{\rm{d}}/\dot{M}(0)$, of $7000-60~000$~yr.} \\
    \end{tabular}}
    \label{bestfit_acc_params_table}
\end{table}

Evident from Figure~\ref{ZnTi_Teff} is that lower initial accretion rates of $\lesssim\!10^{-7}$~\Mdot\ are unable to reproduce the observed depletion of post-AGB binary stars at any effective temperature. In these cases, the depletion timescales are much longer than the star's evolution timescale, which heats up significantly before beginning to become depleted. Further, mildly-depleted post-AGB stars at low effective temperatures ($\lesssim$~5000~K) require both a high initial accretion rate ($\sim\!10^{-6}$~\Mdot) and lower core mass, to be reproduced (see Figure~\ref{ZnTi_Teff}). The slower evolution of the 0.55~\Msun\ models allows more time for the accreted material to be mixed into the outer layers, before the star heats up significantly. We note that since all 0.60~\Msun\ model curves fall below post-AGB star IRAS~17038-4815 in $\left[\rm{Zn}/\rm{Ti}\right]-T_{\rm{eff}}$ space (see Figure~\ref{ZnTi_Teff}), we have taken the closest model curve (red dashed) to be the best-fitting (see Table~\ref{bestfit_acc_params_table}).

In general, initial accretion rates required to reproduce depletion in post-AGB binary stars of current mass 0.55~-~0.60~\Msun\ fall within the range $10^{-7}-10^{-6}$~\Mdot\ (see Figure~\ref{ZnTi_Teff}). Lower disk mass models may only become depleted ($\left[\rm{Zn}/\rm{Ti}\right]>$~0.5~dex) at effective temperatures below 6000~K when paired with a high accretion rate. In the case of faster-evolving 0.60~\Msun\ post-AGB stars, a higher disk mass of $3\times10^{-2}$~\Msun\ is required, in addition to a high accretion rate, to reproduce moderate depletion levels at low temperatures (see lower panel of Figure~\ref{ZnTi_Teff}). When paired with a lower core mass, disk masses of $\gtrsim10^{-2}$~\Msun\ are sufficient to reproduce depletion in most post-AGB stars (see upper panel of Figure~\ref{ZnTi_Teff}).

We further note the degeneracy amongst model curves in Figure~\ref{ZnTi_Teff}, with different initial accretion rate and disk mass combinations. For example, the 0.60~\Msun\ models with $\dot{M}(0)=10^{-6}$~\Mdot\ and $M_{\rm{d}}=7\times10^{-3}$~\Msun\ (red dotted curve), and $\dot{M}(0)=5\times10^{-7}$~\Mdot\ and $M_{\rm{d}}=3\times10^{-2}$~\Msun\ (orange dashed curve), produce approximately the same depletion strength ($[\rm{Zn}/\rm{Ti}]\approx$~3~dex) by the observed effective temperature of HP~Lyr (6300~K), making the `best-fitting' parameters for the object difficult to discern. This is a direct consequence of the inverse proportionality between the two accretion parameters in Equation~\ref{acc_eq}. That is, reproducing depletion in post-AGB stars requires either a low enough initial accretion rate such that disk mass decreases gradually, or a high enough initial disk mass such that there is a larger reservoir available to sustain accretion. Hence, there is no unique combination of disk mass and accretion rate to explain any given observation.

However, our work explores the range of observationally-constrained accretion parameters, which place mass accretion rates and circumbinary disk masses firmly at $\sim10^{-7}$~\Mdot\ and $10^{-2}$~\Msun, respectively. Moreover, we see that mass accretion rates lower than $10^{-7}$~\Mdot\ fail to reproduce observed depletion levels for any given disk mass choice within the observed range. Additionally, mass accretion rates larger than $10^{-6}$~\Mdot\ result in a rapid expansion of the modelled stars, which is non-physical and has no observational-basis in this class of stars. We therefore conclude that despite the degeneracy seen in Figure~\ref{ZnTi_Teff}, our work tests the reasonable range of disk masses and accretion rates.

\subsection{Comparing modelled and observed depletion patterns}

\begin{figure*}[hbt!] % This plot now has NLTE-corrected abundances for EP Lyr.
    \centering
    \includegraphics[width=0.81\linewidth]{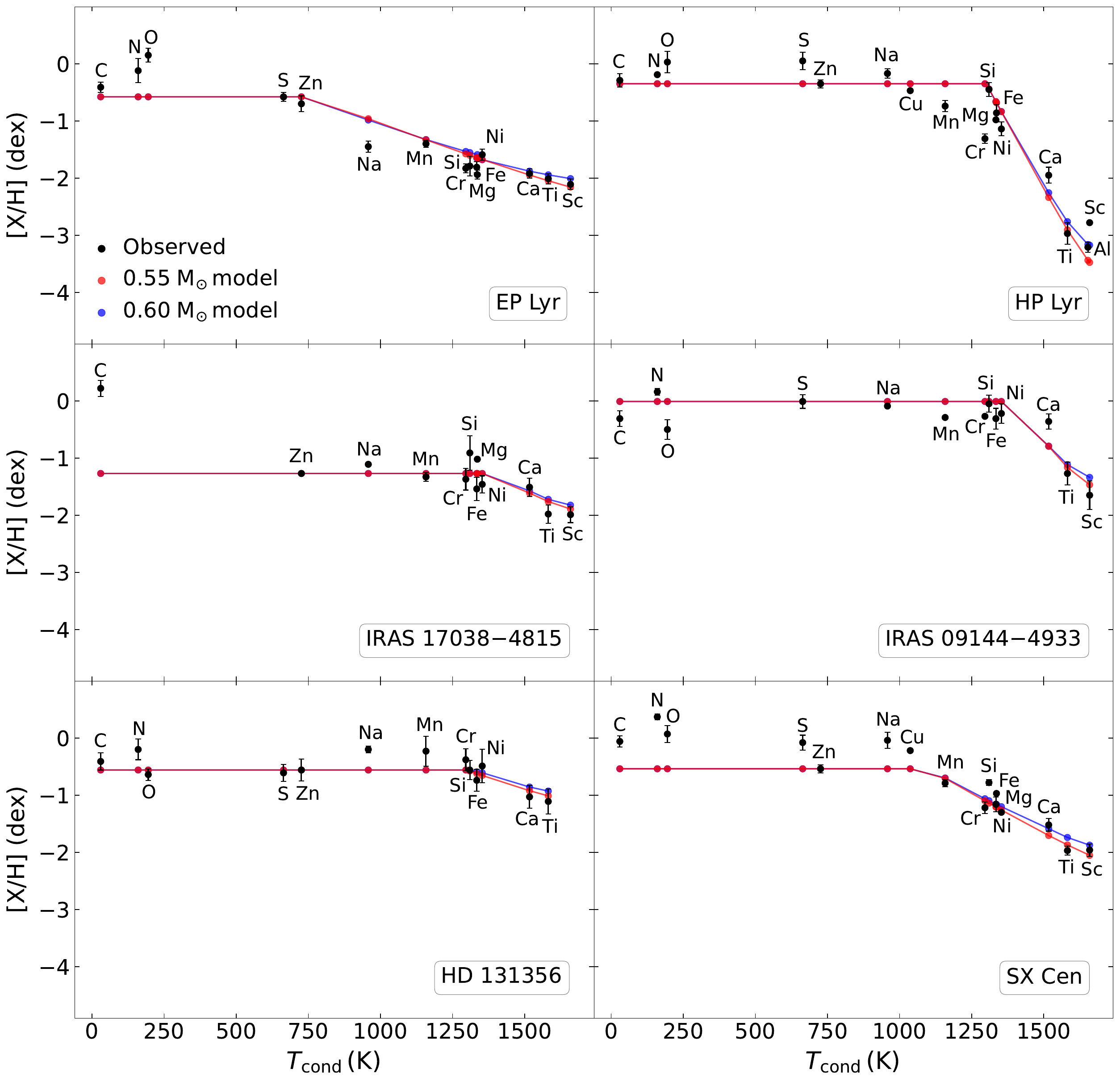}
    \caption{Comparison between observed abundances, and the abundances predicted by the best-fitting 0.55 and 0.60~\Msun\ \mesa\ models from Figure~\ref{ZnTi_Teff}, for chemically-depleted post-AGB stars EP~Lyr ($T_{\rm{eff}}=$~6200~K), HP~Lyr ($T_{\rm{eff}}=$~6300~K), IRAS~17038-4815 ($T_{\rm{eff}}=$~4750~K), IRAS~09144-4933 ($T_{\rm{eff}}=$~5750~K), HD~131356 ($T_{\rm{eff}}=$~6000~K), and SX~Cen ($T_{\rm{eff}}=$~6250~K). Abundances are given relative to solar \protect\cite[values from][]{asplund_2009}, and are presented against condensation temperature, $T_{\rm{cond}}$ \protect\cite[values from][]{lodders_2003}. Predicted model abundances are further normalised to the observed abundances of Zn, or S where Zn is unavailable (i.e., IRAS~09144-4933), as these elements are assumed to be non-depleted and hence representative of the star's initial metallicity.}
    \label{XH_Tcond}
\end{figure*}

To test the depletion modelling process on specific post-AGB binary stars, the object's observed depletion strength and turn-off temperature were set in the model, along with the accretion parameters found to best-fit the star in $\left[\rm{Zn}/\rm{Ti}\right]-T_{\rm{eff}}$ space (see Figure~\ref{ZnTi_Teff} and Table~\ref{bestfit_acc_params_table}). The comparison between observed and predicted model depletion patterns for each star in our sample is presented in Figure~\ref{XH_Tcond}. The predicted model abundances in Figure~\ref{XH_Tcond} were taken at the observed effective temperature of the post-AGB star (see Table~\ref{star_prop_table}), near the surface of the model (where opacity is $<$~1). The predicted abundances have been further normalised to the observed abundance of Zn, or S where Zn is unavailable (i.e., IRAS~09144-4933), as these elements are assumed to be non-depleted and hence representative of initial metallicity \cite[as in][]{Oomen_2019, Oomen_2020}. 
% surface abundances: from "outer edge of outer cell" (mesa docs)
% surface conditions: evaluated at the photosphere (tau_factor < 1) or surface of the model (tau_factor >= 1) when photosphere is not inside the model

Overall, the observed depletion patterns of our post-AGB binary star sample are well-reproduced by the 0.55~\Msun\ models (see Figure~\ref{XH_Tcond}). For cooler ($T_{\rm{eff}}<$~6000~K) post-AGB stars with mild-to-moderate depletion levels (e.g., IRAS~17038-4815 and IRAS~09144-4933), the slower evolution timescale of the 0.55~\Msun\ models allows depletion to occur before the star heats up significantly; the 0.60~\Msun\ models heat too rapidly to be able to reproduce the observed depletion at these lower temperatures. For hotter post-AGB stars with moderate to strong depletion (e.g., EP~Lyr, HP~Lyr, HD~131356, and SX~Cen), the slower evolution of the 0.55~\Msun\ models provides more time for the photosphere to become significantly depleted by the accreted material, while the star evolves to higher effective temperatures; the faster-evolving 0.60~\Msun\ models fail to achieve such depletion levels at high effective temperatures.
 
In general, the best-fitting accretion parameters from Figure~\ref{ZnTi_Teff} provide a good fit to observed abundances, particularly at high condensation temperatures ($T_{\rm{cond}}>$~1400~K). Since model predictions are based on condensation temperatures determined in \textit{ideal} conditions, specifically at solar composition with constant pressure in equilibrium chemistry, we do not expect that all elements are fit exactly by the models. The observed depleted abundances of post-AGB objects reflects the non-ideal nature of circumstellar environments, of which the condensation sequence details are influenced by factors such as metallicity, C/O ratio, and variable pressure. Despite this, the simple assumptions underlying condensation temperature determinations fit very well the observed depletion of post-AGB binary stars, which is remarkable over so many objects.

\section{Results: effects on post-AGB evolution}
\label{evol_effects_section}

%The post-AGB phase begins once a star's envelope is reduced (typically by strong AGB winds, binary interaction, or both) to $\lesssim$~0.02~\Msun. This occurs at an effective temperature of around 4000~K for stars leaving the AGB with masses in the range 0.55~-~0.60~\Msun\ \cite[]{Miller_Bertolami_2016, Kamath_et_al_2023}. Post-AGB evolution continues until the star is hot enough \cite[$\sim$~25~000~K;][]{Schonberner_1987, VanWinckel_2003} to ionise its circumstellar material to form a visible PN. The duration of the post-AGB phase (i.e., the post-AGB evolution timescale, $\mathlarger{\mathlarger{\tau}}_{\rm{\scriptsize{pAGB}}}$) is determined by the rate of further envelope mass removal, which typically takes $10^3-10^4$~years \cite[]{Van_Winckel_2018}. However, accreted gas from a circumbinary disk is found to extend the post-AGB lifetime of a star \cite[]{Oomen_2019}, as the accreted material replenishes the envelope mass, stalling the star at lower effective temperatures. An extended post-AGB phase can have important consequences on PN formation, as the star must reach ionisation temperature within the PN visibility timescale of $\sim$~20~000~-~50~000~years \cite[]{PNE_exp_vel, PN_timescale}, to sufficiently ionise its surroundings before the circumstellar material density decreases excessively. 

The post-AGB phase begins once a star's envelope is reduced to $\lesssim$~0.02~\Msun, typically by strong AGB winds, binary interaction, or both, and continues until the star heats to $\sim$~25~000~K \cite[]{Schonberner_1987, VanWinckel_2003}, at which point it can ionise its circumstellar material to form a visible PN. The duration of the post-AGB phase (i.e., the post-AGB evolution timescale, $\mathlarger{\mathlarger{\tau}}_{\rm{\scriptsize{pAGB}}}$) is determined by the rate of further envelope mass removal, which typically takes $10^3-10^4$~years \cite[]{Van_Winckel_2018}. However, accreted gas from a circumbinary disk can extend the post-AGB lifetime of a star \cite[]{Oomen_2019}, as the accreted material replenishes the envelope mass, stalling the star at lower effective temperatures. An extended post-AGB phase can have important consequences on PN formation, as the star must reach ionisation temperature within the PN visibility timescale of $\sim$~20~000~-~50~000~years \cite[]{PNE_exp_vel, PN_timescale}, to sufficiently ionise its surroundings before the circumstellar material density decreases excessively. 

In this section, we investigate how accretion from a disk impacts important stellar properties, and consequently the duration of the post-AGB phase. Further, we compare the post-AGB evolution timescale with the PN visibility timescale, to infer whether PN formation in the system may be prevented under certain conditions.

\subsection{Stellar structure and properties}
\label{effects_on_properties}

\begin{figure}[hbt!]
    \centering
    \includegraphics[width=0.87\linewidth]{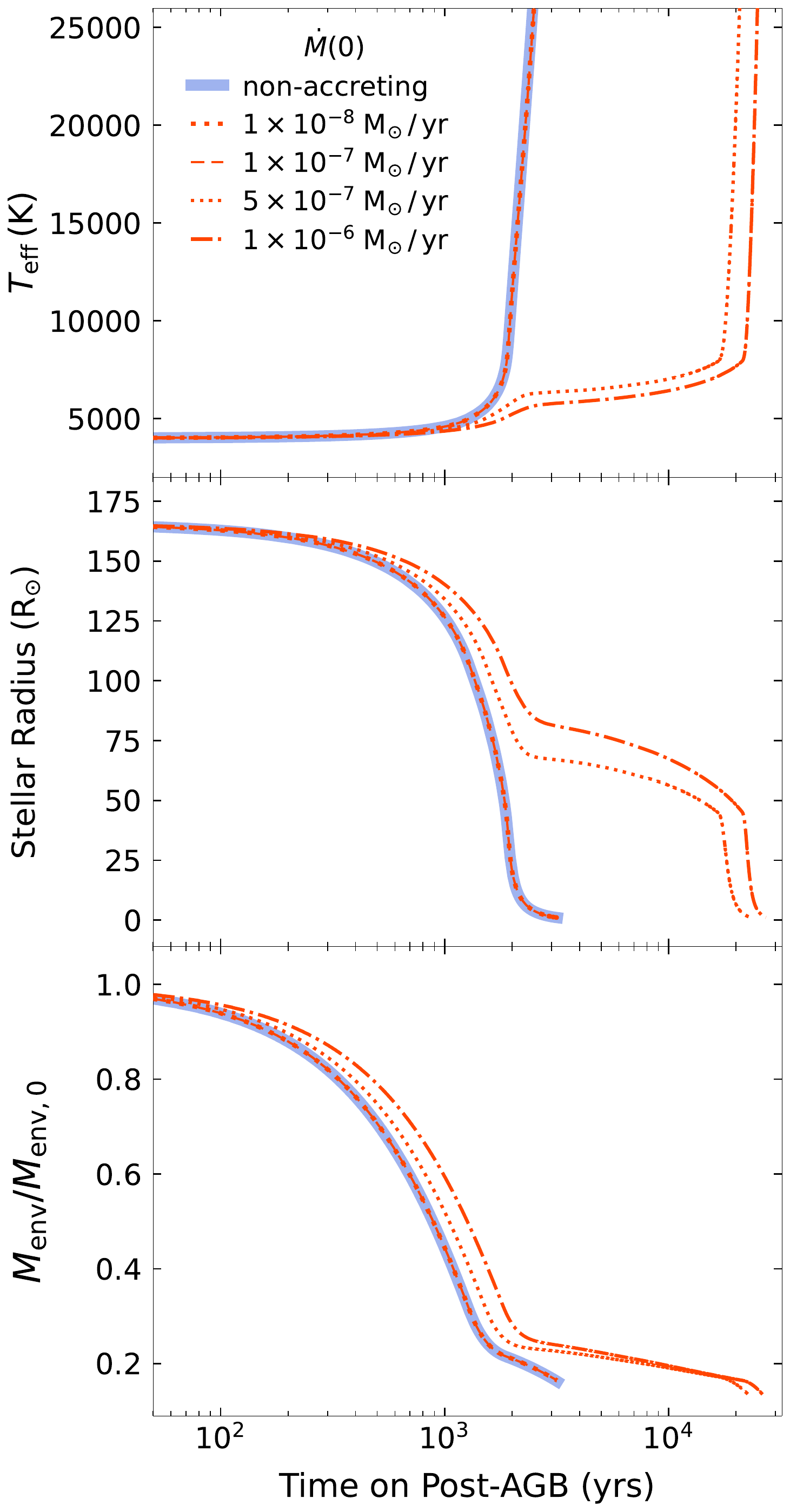}
    \caption{Evolution of effective temperature, stellar radius, and envelope mass ($M_{\rm{env}}$, relative to the start of the post-AGB where $M_{\rm{env},0}\approx$~0.02~\Msun), over the post-AGB phase of the 0.60~\Msun\ models. Accreting models (orange) with different initial accretion rates, $\dot{M}(0)$ (in \Mdot), are indicated by linestyle. A disk mass of $3\times10^{-2}$~\Msun\ was used in all accreting cases. The non-accreting 0.60~\Msun\ model (blue) is shown for reference.}
    \label{all_vs_time_0pt60Msun}
\end{figure}

Figure~\ref{all_vs_time_0pt60Msun} shows the evolution of effective temperature, stellar radius, and hydrogen-rich envelope mass, in the 0.60~\Msun\ models over the post-AGB, for varying initial accretion rates, and an initial disk mass of $3\times10^{-2}$~\Msun. Accretion at high initial rates ($\gtrsim\!5\times10^{-7}$~\Mdot) is shown to significantly impact the effective temperature of post-AGB stars, keeping them cooler for longer and prolonging their post-AGB lifetime. Specifically, effective temperature is seen in Figure~\ref{all_vs_time_0pt60Msun} (upper panel) to be stalled at around 7000~K for approximately 25~000~years. Effective temperature then quickly (within $\sim$~2000~yr) increases to 25~000~K, as by this point the disk mass has been almost entirely exhausted, and the time dependent accretion rate (Equation~\ref{acc_eq}) quickly drops below the rate of nuclear burning at the envelope base ($\sim10^{-7}$~\Mdot). The impact of such rapid accretion on the radius of post-AGB stars is equivalent to that seen for effective temperature; as the star is kept cooler for longer, the natural contraction of the star over the post-AGB is also delayed (see Figure~\ref{all_vs_time_0pt60Msun} middle panel). 

The evolution of hydrogen-rich envelope mass over the post-AGB is additionally presented in Figure~\ref{all_vs_time_0pt60Msun} (lower panel) for the 0.60~\Msun\ models; envelope mass is expressed as a fraction relative to the envelope mass at the start of the post-AGB (i.e., $\sim$~0.02~\Msun). Material accreted from the disk at high initial rates ($\gtrsim5\times10^{-7}$~\Mdot) slows down the envelope removal by temporarily replenishing the star's envelope mass, delaying the post-AGB evolution. At around 20~000~years into the post-AGB phase, the remaining envelope mass is quickly removed by strong stellar winds and nuclear burning, as accretion becomes insufficient to replenish the envelope.

\begin{table*}[hbt!]
    \centering
    \caption{Evolution timescales and extension factors (in square brackets) for the post-AGB phase of the accreting 0.60~\Msun\ models, for each initial accretion rate, $\dot{M}(0)$, and disk mass, $M_{\rm{d}}$, combination.  The post-AGB evolution timescale ($\mathlarger{\mathlarger{\tau}}_{\rm{pAGB}}$) corresponds to the time taken for the model to evolve from $T_{\rm{eff}}=$~4000~K (the start of the post-AGB) to $T_{\rm{eff}}=$~25~000~K (the minimum PN ionisation temperature). The post-AGB phase extension factor was computed as $\mathlarger{\mathlarger{\tau}}_{\rm{pAGB}}/\mathlarger{\mathlarger{\tau}}_{\rm{\scriptsize{norm}}}$, where $\mathlarger{\mathlarger{\tau}}_{\rm{\scriptsize{norm}}}$ is the post-AGB evolution timescale of the non-accreting model (noted below the table for reference).}
    \setlength{\tabcolsep}{4.5pt}
    \renewcommand{\arraystretch}{1.0}
    \resizebox{\linewidth}{!}{\begin{tabular}{p{30mm} >{\centering\arraybackslash}p{34mm} >{\centering\arraybackslash}p{34mm} >{\centering\arraybackslash}p{34mm} >{\centering\arraybackslash}p{37mm}}
         \hline
         \hline
         \rule{0pt}{3mm}$\dot{M}(0)$ (\,\Mdot\,) & $1\times10^{-8}$ & $1\times10^{-7}$ & $5\times10^{-7}$ & $1\times10^{-6}$ \\[0.5ex]
         \hline
         %\\[-4ex]
         $M_{\rm{d}} \,=\, 3\times10^{-2}\,$~\Msun   &  $2471~\rm{yr}$ \;\;$\left[\,1.01\,\right]$  &  $2723~\rm{yr}$ \;\;$\left[\,1.12\,\right]$  &  $20~601~\rm{yr}$ \;\;$\left[\,8.44\,\right]$  &  \;\;$24~793~\rm{yr}$ \;\;$\left[\,10.16\,\right]$   \\
         $M_{\rm{d}} \,=\, 1\times10^{-2}\,$~\Msun   &  $2471~\rm{yr}$ \;\;$\left[\,1.01\,\right]$  &  $2670~\rm{yr}$ \;\;$\left[\,1.09\,\right]$  &  \;\;\;$9675~\rm{yr}$ \;\;$\left[\,3.96\,\right]$  &  $\;\;10~733~\rm{yr} \;\; \;\;\left[\,4.40\,\right]$    \\
         $M_{\rm{d}} \,=\, 7\times10^{-3}\,$~\Msun   &  $2471~\rm{yr}$ \;\;$\left[\,1.01\,\right]$  &  $2471~\rm{yr}$ \;\;$\left[\,1.01\,\right]$  &  \;\;\;$7627~\rm{yr}$ \;\;$\left[\,3.13\,\right]$  &  $\;\;\;8253~\rm{yr} \;\;\; \;\;\,\left[\,3.38\,\right]$ \\ [0.5ex]
         \hline
         \multicolumn{5}{p{\linewidth}}{\textbf{Note.} The non-accreting $0.60$~\Msun\ model has $\mathlarger{\mathlarger{\tau}}_{\rm{\scriptsize{norm}}}=2440$~years.} \\
    \end{tabular}}
    \label{PAGB_ext_factors_results_table}
\end{table*}

\subsection{Evolution timescale}

The stalling of post-AGB stars at lower temperatures as a result of accretion (see Section~\ref{effects_on_properties}) effectively extends their post-AGB evolution timescale. In Table~\ref{PAGB_ext_factors_results_table}, we present the evolution timescale and extension factor for the post-AGB phase of the accreting 0.60~\Msun\ models, for different combinations of initial accretion rate and disk mass. The post-AGB evolution timescale, $\mathlarger{\mathlarger{\tau}}_{\rm{pAGB}}$, was computed as the time taken for the model to evolve from the start of the post-AGB (i.e., taken to be $T_{\rm{eff}}=$~4000~K in this work), to an effective temperature of 25~000~K. The extension factor for the post-AGB phase was then computed as $\mathlarger{\mathlarger{\tau}}_{\rm{pAGB}}/\mathlarger{\mathlarger{\tau}}_{\rm{norm}}$, where $\mathlarger{\mathlarger{\tau}}_{\rm{norm}}$ is the post-AGB evolution timescale of the model with no accretion applied. Specifically, the non-accreting 0.60~\Msun\ model was found to have a post-AGB evolution timescale of only 2440~years.

The general trend arising from Table~\ref{PAGB_ext_factors_results_table} is that more rapid accretion from a disk of higher mass has the strongest impact on post-AGB evolution (as was seen in Figure~\ref{all_vs_time_0pt60Msun}). Initial accretion rates of $\gtrsim5\times10^{-7}$~\Mdot\ are required to significantly extend the post-AGB phase of a star. Lower initial accretion rates ($\lesssim10^{-7}$~\Mdot) have little to no impact on post-AGB evolution, resulting in small extension factors of up to only 1.12 (see Table~\ref{PAGB_ext_factors_results_table}). Initial disk mass has a substantial impact on the post-AGB evolution timescales of the more rapidly accreting ($\gtrsim\!5\times10^{-7}$~\Mdot) models, which show significantly extended post-AGB phases by factors of $\sim$~3~-~10 (see Table~\ref{PAGB_ext_factors_results_table}).

\subsubsection{Consequences for PN formation}

For a star to ionise its circumstellar material and form a visible PN, it must evolve to an effective temperature of $\sim$~25~000~K within the PN visibility timescale of 20~000~-~50~000~years \cite[]{PNE_exp_vel, PN_timescale}. The specific PN visibility timescale is governed by the stellar and circumstellar properties, which influence how rapidly the surrounding material expands away. The distance the circumstellar material has moved from the star by the time it reaches $\sim$~25~000~K determines the ionisation strength of the resulting PN. To ensure the formation of a PN, the ionisation much hence occur on a shorter timescale than the PN visibility timescale.

To determine whether accretion may delay the evolution of post-AGB binary stars sufficiently enough to avoid forming a visible PN, we compared the computed post-AGB evolution timescales of the 0.60~\Msun\ stellar models in Table~\ref{PAGB_ext_factors_results_table} to the PN visibility timescale of 20~000~-~50~000~years. Based on Table~\ref{PAGB_ext_factors_results_table}, almost all models have post-AGB evolution timescales much shorter than the minimum PN visibility timescale of 20~000~years, implying they are capable of forming ionised PNe given favourable conditions. Only two models, which are experiencing rapid accretion ($\gtrsim5\times10^{-7}$~\Mdot) from a disk of high mass ($3\times10^{-2}$~\Msun), present post-AGB evolution timescales of $\sim$~20~000~-~25~000~years (see Table~\ref{PAGB_ext_factors_results_table}), slightly exceeding the minimum PN visibility timescale.

\begin{figure}[hbt!]
    \centering
    \includegraphics[width=\linewidth]{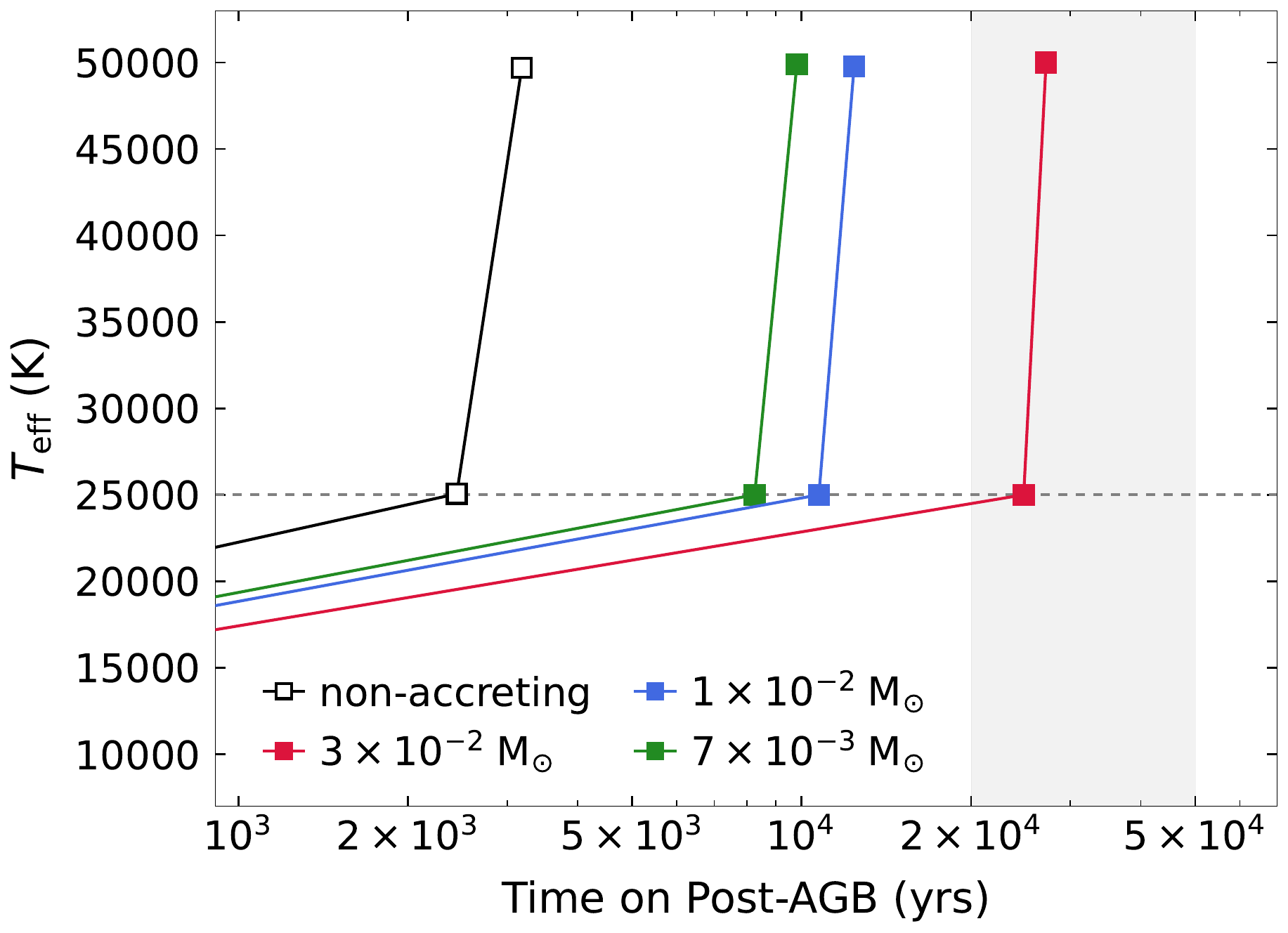}
    \caption{Post-AGB effective temperature evolution for the non-accreting 0.60~\Msun\ model and most rapidly accreting ($\dot{M}(0)=10^{-6}$~\Mdot) 0.60~\Msun\ models of different disk mass (as indicated). Effective temperature increases from the start of the post-AGB to the points marked at 25~000~K (the minimum PN ionisation temperature, indicated by a dashed horizontal line) and 50~000~K. The shaded region between 20~000 and 50~000 years corresponds to the range of PN visibility timescales; post-AGB stars that reach $T_{\rm{eff}}\sim$~25~000~K after this time are unlikely to form a visible PN.}
    \label{Teff_vs_time_PN_formation}
\end{figure}

Figure~\ref{Teff_vs_time_PN_formation} visualises this comparison for the non-accreting 0.60~\Msun\ model, and most rapidly accreting 0.60~\Msun\ model with $\dot{M}(0)=10^{-6}$~\Mdot\ for different disk mass values. Models with disk mass $\lesssim10^{-2}$~\Msun\ reach 25~000~K by $\sim$~10~000~years into the post-AGB, which is much shorter than the PN visibility timescale. The post-AGB evolution timescale of the model with the highest disk mass ($3\times10^{-2}$~\Msun) is shown to slightly exceed the minimum PN visibility timescale of 20~000~years (see Figure~\ref{Teff_vs_time_PN_formation}). However, the model still reaches 25~000~K well within the upper limit on the predicted PN visibility timescale (i.e., 50~000~yr). For systems with very short visibility timescales (closer to 20~000~yr), such a small extension of the post-AGB phase may be enough to avoid the formation of an ionised PN. However, it is more likely that the star will be hot enough in time to ionise a slightly more diffuse, yet still close circumstellar material. 

%the minimum PN ionisation temperature of $25~000$~K is marked by a dashed horizontal line. 

Since all models are found to reach the minimum ionisation temperature within, or close to the minimum PN visibility timescale of 20~000~years, we hence conclude that the avoidance of a PN in our modelled systems is unlikely. 

\section{Discussion}

We have used the one-dimensional stellar evolution code \mesa\ to model accretion onto post-AGB stars, with the assumption that the accretion flow originates from a circumbinary disk depleted in refractory elements. The aim was to reproduce the chemical depletion patterns observed in six specific post-AGB binary stars, namely EP~Lyr, HP~Lyr, IRAS~17038-4815, IRAS~09144-4933, HD~131356, and SX~Cen. Our findings show that gas rapidly accreted onto the surface of post-AGB stars significantly impacts post-AGB evolution, by slowing down the rate of envelope mass removal and keeping the star cooler for longer. However, we find that in most cases, an extended post-AGB evolution timescale is insufficient to prevent the formation of ionised PNe in post-AGB binary systems.

We noted that starting accretion at the onset of the post-AGB can maintain the large radius of the star which has only just departed the AGB, potentially causing a temporary growth beyond the Roche lobe (RL) that triggers episodic mass transfer. Additionally, some stars may still be partly filling their RL upon entering the post-AGB, leading to continued mass transfer. From our sample, post-AGB binary star IRAS~17038-4815 was estimated to be larger than its RL radius at periastron (see Table~\ref{star_prop_table}). While additional mass-transfer could significantly impact surface chemistry, envelope mass, and other stellar properties, the wide orbital separations ($>$~1000~days) of most of our post-AGB systems suggest such effects would have minimal impact on the results of our study. Additionally, the late thermal pulse (LTP) observed in the 0.55~\Msun\ models (see \ref{Appendix_LTP}) suggests that rapid mass accretion, which extends the post-AGB phase, may increase the likelihood of such phenomenon for stars that leave the AGB during a thermal pulse cycle, by providing more time for the LTP to occur. Given these considerations, we present the main findings of our work, and their implications, below.

\subsection{Chemical depletion and accretion rates}

Overall, our results show that initial accretion rates of $\sim\!10^{-7}-10^{-6}$~\Mdot\ are required to reproduce observed chemical depletion levels within short post-AGB timescales. This is in agreement with the mass accretion rates reported by previous theoretical studies of chemical depletion \cite[e.g.,][]{Oomen_2019, Oomen_2020}, and is consistent with observationally-derived mass-accretion rates from jet absorption features \cite[i.e., $10^{-7}-10^{-5}$~\Mdot; see][]{bollen_2017, Bollen_2021, Bollen_2022, ToonsPaper_2024}. The high accretion rates required to reproduce the observed chemical depletion of post-AGB binary stars allow us to infer the longevity of circumbinary disks, pointing to disk lifetimes of $\sim$~10~000~-~100~000~years.

The predicted, depleted abundances of some elements (e.g., Na, Mn, Cr, and Mg) were often either over- or under-estimated by our models. However, we concluded that the abundance estimation errors on these elements, in many cases, placed them closer to model predictions. For EP~Lyr, NLTE-corrected abundances recently published in \cite{Maksym_2023b} provided depletion pattern abundance estimations with significantly reduced observational errors. Further, the linear relation predicted by our models is based on condensation temperature determinations valid only in ideal conditions, which do not exist in the complex circumstellar environments of post-AGB binary stars with dusty circumbinary disks; the condensation sequence of post-AGB stars in nature is influenced by additional factors, such as metallicity, C/O ratio, and variable pressure, meaning that predicted model abundances cannot be expected to fit observed depleted abundances exactly.

The degeneracy amongst model curves with different accretion rates and disk masses (Figure~\ref{ZnTi_Teff}) makes the identification of a unique set of accretion parameters required to reproduce observed depletion levels challenging. That is, models with higher accretion rates and lower disk masses yield similar depletion strengths to models with lower accretion rates and higher disk masses, at certain effective temperatures. That said, values of disk mass and mass accretion rates outside of the tested range are not supported by observations. Moreover, lower accretion rates completely fail to generate a depleted abundance pattern for the star, while higher accretion rates would dictate an immediate inflation of the star back to the AGB. We therefore conclude that reproducing depletion in IRAS~17038-4815 and IRAS~09144-4933 requires the highest accretion rate with the most massive disk; EP~Lyr, HP~Lyr, HD~131356, and SX~Cen, may be reasonably reproduced by accretion rates between $10^{-7}-5\times10^{-7}$~\Mdot, for any given disk mass tested here, making these objects quite degenerate.

\subsection{Implications for post-AGB evolution and PN formation}

Accretion at the high rates required to reproduce observed chemical depletion was found to efficiently stall the effective temperature evolution, and hence radial contraction, of a star over the post-AGB phase. Such rapidly accreted material is sufficient to temporarily replenish the stellar envelope mass, keeping the star cooler for longer. Disk mass was shown to have a large impact on the post-AGB lifetime, as high accretion rates can be sustained for longer. Specifically, post-AGB evolution timescales were shown to be significantly extended by up to a factor of ten. However, all post-AGB evolution timescales were still much shorter than the PN visibility timescale. In the most extended case, the post-AGB evolution timescale exceeded the minimum PN visibility timescale (i.e., 20~000~yr) by only $\sim$~5000~years, still being much shorter than the upper limit on the PN visibility timescale (i.e., 50~000~yr).

\subsection{Concluding remarks}

%An extension of the post-AGB phase by rapid accretion is insufficient to avoid the ionisation of circumstellar material in these systems. However, the post-AGB evolution timescale is strongly dependent on core mass, with lower mass post-AGB stars presenting much longer evolution timescales; under the right circumstances, post-AGB phase extension by accretion \textit{may} be significant enough to avoid the formation of a PN in these systems. Additionally, stellar winds during the post-AGB phase, of which the details remain almost completely unknown, impact the rate of envelope mass removal from the star, and hence its post-AGB lifetime. An accurate assessment of PN formation in post-AGB systems hence requires the additional consideration of many factors beyond the scope of this study, including detailed information on the circumstellar material, and stellar and binary properties. Regardless, our results provide a reasonable indication of the plausibility of PN formation in post-AGB binary systems, given typical, observationally-constrained accretion parameter values. 

High accretion rates ($\sim\!10^{-7}-10^{-6}$~\Mdot) for typical disk masses ($\sim10^{-2}$~\Msun) are required to reproduce chemical depletion within short post-AGB timescales. While this rapid accretion extends the post-AGB phase, it is insufficient to prevent the ionisation of circumstellar material in these systems. However, post-AGB evolution timescales depend strongly on core mass, with lower mass post-AGB stars presenting much longer evolution timescales; under the right circumstances, post-AGB phase extension by accretion \textit{may} be significant enough to avoid the formation of a PN. Additionally, stellar winds during the post-AGB phase, of which the details remain almost completely unknown, impact the rate of envelope mass removal from the star, and hence its post-AGB lifetime. An accurate assessment of PN formation in post-AGB systems hence requires additional consideration of factors beyond the scope of this study, including detailed information on the circumstellar material, and stellar and binary properties. Regardless, our results provide a reasonable indication of the plausibility of PN formation in post-AGB binary systems, given typical, observationally-constrained accretion parameter values.

\begin{acknowledgement}
KM acknowledges the financial support provided by the Macquarie Research Excellence Scholarship (MQRES) program, received throughout the duration of this research. DK, ODM and HVW acknowledge the support from the Australian Research Council Discovery Project DP240101150. Further, we thank the anonymous referee for their comments, which helped to improve the quality and clarity of this work.
\end{acknowledgement}

%\paragraph{Funding Statement}
%This research was supported by grants from the <funder-name> <doi> (<award ID>); <funder-name> <doi> (<award ID>).

%\paragraph{Competing Interests}
%A statement about any financial, professional, contractual or personal relationships or situations that could be perceived to impact the presentation of the work --- or `None' if none exist.

\paragraph{Data Availability Statement}
The \mesa\ models used in this work have been made publicly available on GitHub at \href{https://github.com/Kayla101197/MESA_Accreting_2.5Msun.git}{this link}. This repository contains the \mesa\ inlists and other source files used in our study, and provides an outline of any modifications made to the code to allow simultaneous mass gain by accreted gas and mass loss by stellar winds.

%%%%% BIBLIOGRAPHY %%%%%
\printendnotes

%\printbibliography
%\bibliographystyle{paslike}
\bibliography{main}

\begin{thebibliography}{}
\expandafter\ifx\csname natexlab\endcsname\relax\def\natexlab#1{#1}\fi

\bibitem[{{Andrych} {et~al.}(2023){Andrych}, {Kamath}, {Kluska}, {Van Winckel}, {Ertel}, \& {Corporaal}}]{Andrych_et_al_2023}
{Andrych}, K., {Kamath}, D., {Kluska}, J., {et~al.} 2023, MNRAS, 524, 4168

\bibitem[{{Andrych} {et~al.}(2024){Andrych}, {Kamath}, {Van Winckel}, {Kluska}, {Schmid}, {Corporaal}, \& {Milli}}]{Katya_paper_2024}
{Andrych}, K., {Kamath}, D., {Van Winckel}, H., {et~al.} 2024, MNRAS, 535, 1763

\bibitem[{{Annibali} {et~al.}(2017){Annibali}, {Tosi}, {Romano}, {Buzzoni}, {Cusano}, {Fumana}, {Marchetti}, {Mignoli}, {Pasquali}, \& {Aloisi}}]{PN_timescale}
{Annibali}, F., {Tosi}, M., {Romano}, D., {et~al.} 2017, ApJ, 843, 20

\bibitem[{Asplund {et~al.}(2009)Asplund, Grevesse, Sauval, \& Scott}]{asplund_2009}
Asplund, M., Grevesse, N., Sauval, A.~J., \& Scott, P. 2009, ARA\&A, 47, 481

\bibitem[{{Bailer-Jones} {et~al.}(2021){Bailer-Jones}, {Rybizki}, {Fouesneau}, {Demleitner}, \& {Andrae}}]{GaiaDR3_dist_BailerJones2021}
{Bailer-Jones}, C.~A.~L., {Rybizki}, J., {Fouesneau}, M., {Demleitner}, M., \& {Andrae}, R. 2021, {VizieR Online Data Catalog: Distances to 1.47 billion stars in Gaia EDR3 (Bailer-Jones+, 2021)}, VizieR On-line Data Catalog: I/352. Originally published in: 2021AJ....161..147B

\bibitem[{{Bl\"{o}cker}(1995)}]{Blocker}
{Bl\"{o}cker}, T. 1995, A\&A, 297, 727

\bibitem[{Bl\"{o}cker(2001)}]{Blocker_2001}
Bl\"{o}cker, T. 2001, Ap\&SS, 275, 1–14

\bibitem[{{Bollen} {et~al.}(2022){Bollen}, {Kamath}, {Van Winckel}, {De Marco}, {Verhamme}, {Kluska}, \& {Wardle}}]{Bollen_2022}
{Bollen}, D., {Kamath}, D., {Van Winckel}, H., {et~al.} 2022, A\&A, 666, A40

\bibitem[{{Bollen} {et~al.}(2021){Bollen}, {Kamath}, {Van Winckel}, {De Marco}, \& {Wardle}}]{Bollen_2021}
{Bollen}, D., {Kamath}, D., {Van Winckel}, H., {De Marco}, O., \& {Wardle}, M. 2021, MNRAS, 502, 445

\bibitem[{{Bollen} {et~al.}(2017){Bollen}, {Van Winckel}, \& {Kamath}}]{bollen_2017}
{Bollen}, D., {Van Winckel}, H., \& {Kamath}, D. 2017, A\&A, 607, A60

\bibitem[{{Bujarrabal} {et~al.}(2013){Bujarrabal}, {Alcolea}, {Van Winckel}, {Santander-Garc\'{\i}a}, \& {Castro-Carrizo}}]{Bujarrabal_2013}
{Bujarrabal}, V., {Alcolea}, J., {Van Winckel}, H., {Santander-Garc\'{\i}a}, M., \& {Castro-Carrizo}, A. 2013, A\&A, 557, A104

\bibitem[{{Bujarrabal} {et~al.}(2015){Bujarrabal}, {Castro-Carrizo}, {Alcolea}, \& {Van Winckel}}]{Bujarrabal_et_al_2015}
{Bujarrabal}, V., {Castro-Carrizo}, A., {Alcolea}, J., \& {Van Winckel}, H. 2015, A\&A, 575, L7

\bibitem[{{Bujarrabal} {et~al.}(2017){Bujarrabal}, {Castro-Carrizo}, {Alcolea}, {Van Winckel}, {Contreras}, \& {Santander-Garc{\'\i}a}}]{bujarrabal_2017}
{Bujarrabal}, V., {Castro-Carrizo}, A., {Alcolea}, J., {et~al.} 2017, A\&A, 597, L5

\bibitem[{{Bujarrabal} {et~al.}(2018){Bujarrabal}, {Castro-Carrizo}, {Van Winckel}, {Alcolea}, {S\'anchez Contreras}, {Santander-Garc\'{\i}a}, \& {Hillen}}]{Bujarrabal_et_al_2018}
{Bujarrabal}, V., {Castro-Carrizo}, A., {Van Winckel}, H., {et~al.} 2018, A\&A, 614, A58

\bibitem[{{Bujarrabal} {et~al.}(2007){Bujarrabal}, {Van Winckel}, {Neri}, {Alcolea}, {Castro-Carrizo}, \& {Deroo}}]{Bujarrabal_et_al_2007}
{Bujarrabal}, V., {Van Winckel}, H., {Neri}, R., {et~al.} 2007, A\&A, 468, L45

\bibitem[{{Corporaal} {et~al.}(2023{\natexlab{a}}){Corporaal}, {Kluska}, {Van Winckel}, {Andrych}, {Cuello}, {Kamath}, \& {M\'erand}}]{Corporaal_2023b}
{Corporaal}, A., {Kluska}, J., {Van Winckel}, H., {et~al.} 2023{\natexlab{a}}, A\&A, 674, A151

\bibitem[{{Corporaal} {et~al.}(2023{\natexlab{b}}){Corporaal}, {Kluska}, {Van Winckel}, {Kamath}, \& {Min}}]{corporaal_2023a}
{Corporaal}, A., {Kluska}, J., {Van Winckel}, H., {Kamath}, D., \& {Min}, M. 2023{\natexlab{b}}, A\&A, 671, doi:10.1051/0004-6361/202245689

\bibitem[{Cranmer \& Saar(2011)}]{Cranmer_2011}
Cranmer, S.~R., \& Saar, S.~H. 2011, ApJ, 741, 54

\bibitem[{{De Prins} {et~al.}(2024){De Prins}, {Van Winckel}, {Ferreira}, {Verhamme}, {Kamath}, {Zimniak}, \& {Jacquemin-Ide}}]{ToonsPaper_2024}
{De Prins}, T., {Van Winckel}, H., {Ferreira}, J., {et~al.} 2024, A\&A, 689, A151

\bibitem[{{Dell’Agli} {et~al.}(2021){Dell’Agli}, {Marini}, {D’Antona}, {Ventura}, {Groenewegen}, {Mattsson}, {Kamath}, {García-Hernández}, \& {Tailo}}]{Dellagli_et_al_2021}
{Dell’Agli}, F., {Marini}, E., {D’Antona}, F., {et~al.} 2021, MNRAS Lett., 502, L35

\bibitem[{{Eggleton}(1983)}]{eggleton_1983}
{Eggleton}, P. 1983, ApJ, 268, 368

\bibitem[{{Gallardo Cava} {et~al.}(2021){Gallardo Cava}, {G\'omez-Garrido}, {Bujarrabal}, {Castro-Carrizo}, {Alcolea}, \& {Van Winckel}}]{Gallardo_et_al_2021}
{Gallardo Cava}, I., {G\'omez-Garrido}, M., {Bujarrabal}, V., {et~al.} 2021, A\&A, 648, A93

\bibitem[{{Gesicki} {et~al.}(2018){Gesicki}, {Zijlstra}, \& {Miller Bertolami}}]{PN_visibility_2018}
{Gesicki}, K., {Zijlstra}, A.~A., \& {Miller Bertolami}, M.~M. 2018, Nature Astronomy, 2, 580

\bibitem[{{Gezer} {et~al.}(2015){Gezer}, {Van Winckel}, {Bozkurt}, {De Smedt}, {Kamath}, {Hillen}, \& {Manick}}]{gezer_et_al_2015}
{Gezer}, I., {Van Winckel}, H., {Bozkurt}, Z., {et~al.} 2015, MNRAS, 453, 133

\bibitem[{{Giridhar} {et~al.}(2005){Giridhar}, {Lambert}, {Reddy}, {Gonzalez}, \& {Yong}}]{Giridhar_2005}
{Giridhar}, S., {Lambert}, D.~L., {Reddy}, B.~E., {Gonzalez}, G., \& {Yong}, D. 2005, ApJ, 627, 432

\bibitem[{{Gonzalez} {et~al.}(1997){Gonzalez}, {Lambert}, \& {Giridhar}}]{Gonzalez_1997}
{Gonzalez}, G., {Lambert}, D.~L., \& {Giridhar}, S. 1997, ApJ, 479, 427

\bibitem[{{Herwig}(2001)}]{Herwig_2001}
{Herwig}, F. 2001, Ap\&SS, 275, 15

\bibitem[{Hillen {et~al.}(2016)Hillen, Kluska, Le~Bouquin, Van~Winckel, Berger, Kamath, \& Bujarrabal}]{hillen_2016}
Hillen, M., Kluska, J., Le~Bouquin, J.-B., {et~al.} 2016, A\&A, 588, L1

\bibitem[{Hillen {et~al.}(2017)Hillen, Van~Winckel, {Menu, J.}, {Manick, R.}, {Debosscher, J.}, {Min, M.}, {de Wit, W.-J.}, {Verhoelst, T.}, {Kamath, D.}, \& {Waters, L. B. F. M.}}]{Hillen_2017}
Hillen, M., Van~Winckel, H., {Menu, J.}, {et~al.} 2017, A\&A, 599, A41

\bibitem[{{Iben} {et~al.}(1983){Iben}, {Kaler}, {Truran}, \& {Renzini}}]{Iben_et_al_1983}
{Iben}, Jr., I., {Kaler}, J.~B., {Truran}, J.~W., \& {Renzini}, A. 1983, ApJ, 264, 605

\bibitem[{{Izzard} \& {Jermyn}(2018)}]{Izzard_Jermyn_2018}
{Izzard}, R., \& {Jermyn}, A. 2018, Galaxies, 6, doi:10.3390/galaxies6030097

\bibitem[{{Jacob} {et~al.}(2013){Jacob}, {Sch\"onberner}, \& {Steffen}}]{PNE_exp_vel}
{Jacob}, R., {Sch\"onberner}, D., \& {Steffen}, M. 2013, A\&A, 558, A78

\bibitem[{Jermyn {et~al.}(2023)Jermyn, Bauer, Schwab, Farmer, Ball, Bellinger, Dotter, Joyce, Marchant, Mombarg, Wolf, Wong, Cinquegrana, Farrell, Smolec, Thoul, Cantiello, Herwig, Toloza, Bildsten, Townsend, \& Timmes}]{MESA_update_2023}
Jermyn, A.~S., Bauer, E.~B., Schwab, J., {et~al.} 2023, ApJS, 265, 15

\bibitem[{{Joyce} \& {Tayar}(2023)}]{Joyce_and_Tayar_2023}
{Joyce}, M., \& {Tayar}, J. 2023, Galaxies, 11, doi:10.3390/galaxies11030075

\bibitem[{{Kamath} {et~al.}(2023){Kamath}, {Dell’Agli}, {Ventura}, {Van Winckel}, {Tosi}, \& {Karakas}}]{Kamath_et_al_2023}
{Kamath}, D., {Dell’Agli}, F., {Ventura}, P., {et~al.} 2023, MNRAS, 519, 2169

\bibitem[{{Kamath} \& {Van Winckel}(2019)}]{Kamath_VanWinckel_2019}
{Kamath}, D., \& {Van Winckel}, H. 2019, MNRAS, 486, 3524

\bibitem[{{Kamath} {et~al.}(2015){Kamath}, {Wood}, \& {Van Winckel}}]{Kamath_2015}
{Kamath}, D., {Wood}, P.~R., \& {Van Winckel}, H. 2015, MNRAS, 454, 1468

\bibitem[{{Kamath} {et~al.}(2014){Kamath}, {Wood}, {Van Winckel}, \& {author}}]{Kamath_2014}
{Kamath}, D., {Wood}, P.~R., {Van Winckel}, H., \& {author}. 2014, MNRAS, 439, 2211

\bibitem[{{Kamath} {et~al.}(2016){Kamath}, {Wood}, {Van Winckel}, \& {Nie}}]{Kamath_et_al_2016}
{Kamath}, D., {Wood}, P.~R., {Van Winckel}, H., \& {Nie}, J.~D. 2016, A\&A, 586, L5

\bibitem[{{Kluska} {et~al.}(2018){Kluska}, {Hillen}, {Van Winckel}, {Manick}, {Min}, {Regibo}, \& {Royer}}]{Kluska_et_al_2018}
{Kluska}, J., {Hillen}, M., {Van Winckel}, H., {et~al.} 2018, A\&A, 616, A153

\bibitem[{{Kluska} {et~al.}(2022){Kluska}, {Van Winckel}, {Copp\'ee}, {Oomen}, {Dsilva, K.}, {Kamath, D.}, {Bujarrabal, V.}, \& {Min, M.}}]{Kluska_2022}
{Kluska}, J., {Van Winckel}, H., {Copp\'ee}, Q., {et~al.} 2022, A\&A, 658, A36

\bibitem[{{Kluska} {et~al.}(2019){Kluska}, {Van Winckel}, {Hillen}, {Berger}, {Kamath}, {Le Bouquin}, \& {Min}}]{Kluska_2019}
{Kluska}, J., {Van Winckel}, H., {Hillen}, M., {et~al.} 2019, A\&A, 631, A108

\bibitem[{{Kobayashi} {et~al.}(2020){Kobayashi}, {Karakas}, \& {Lugaro}}]{Kobayashi_2020}
{Kobayashi}, C., {Karakas}, A.~I., \& {Lugaro}, M. 2020, \apj, 900, 179

\bibitem[{Lodders(2003)}]{lodders_2003}
Lodders, K. 2003, ApJ, 591, 1220

\bibitem[{{Maas} {et~al.}(2005){Maas}, {Van Winckel}, \& {Lloyd Evans}}]{Maas_et_al_2005}
{Maas}, T., {Van Winckel}, H., \& {Lloyd Evans}, T. 2005, A\&A, 429, 297

\bibitem[{{Maas} {et~al.}(2002){Maas}, {Van Winckel}, \& {Waelkens}}]{Maas_et_al_2002}
{Maas}, T., {Van Winckel}, H., \& {Waelkens}, C. 2002, A\&A, 386, 504

\bibitem[{{Manick} {et~al.}(2021){Manick}, {Miszalski}, {Kamath}, {Whitelock}, {Van Winckel}, {Hrivnak}, {Barlow}, \& {Mohamed}}]{Manick_2021}
{Manick}, R., {Miszalski}, B., {Kamath}, D., {et~al.} 2021, \mnras, 508, 2226

\bibitem[{{Manick} {et~al.}(2017){Manick}, Van~Winckel, Kamath, Hillen, \& Escorza}]{manick_2017}
{Manick}, R., Van~Winckel, H., Kamath, D., Hillen, M., \& Escorza, A. 2017, A\&A, 597, A129

\bibitem[{{Menon} {et~al.}(2024){Menon}, {Kamath}, {Mohorian}, {Van Winckel}, \& {Ventura}}]{Meghna_2024}
{Menon}, M., {Kamath}, D., {Mohorian}, M., {Van Winckel}, H., \& {Ventura}, P. 2024, PASA, 41, e025

\bibitem[{{Miller Bertolami}(2016)}]{Miller_Bertolami_2016}
{Miller Bertolami}, M. 2016, A\&A, 588, A25

\bibitem[{{Miszalski} {et~al.}(2018){Miszalski}, {Manick}, {Mikołajewska}, {Van Winckel}, \& {Iłkiewicz}}]{MyCin18}
{Miszalski}, B., {Manick}, R., {Mikołajewska}, J., {Van Winckel}, H., \& {Iłkiewicz}, K. 2018, PASA, 35, e027

\bibitem[{{Mohorian} {et~al.}(2025){Mohorian}, {Kamath}, {Menon}, {Amarsi}, {Van Winckel}, {Fava}, \& {Andrych}}]{Maksym_2023b}
{Mohorian}, M., {Kamath}, D., {Menon}, M., {et~al.} 2025, MNRAS, staf375

\bibitem[{Mohorian {et~al.}(2024)Mohorian, {Kamath}, {Menon}, {Ventura}, {Van Winckel}, {Garc{\'\i}a-Hern{\'a}ndez}, \& {Masseron}}]{Maksym_2023a}
Mohorian, M., {Kamath}, D., {Menon}, M., {et~al.} 2024, MNRAS, 530, 761

\bibitem[{{M{\"o}sta} {et~al.}(2019){M{\"o}sta}, {Taam}, \& {Duffell}}]{Mosta_et_al_2019}
{M{\"o}sta}, P., {Taam}, R., \& {Duffell}, P. 2019, ApJ, 875, L21

\bibitem[{{Mu{\~n}oz} {et~al.}(2019){Mu{\~n}oz}, {Miranda}, \& {Lai}}]{Munoz_et_al_2019}
{Mu{\~n}oz}, D., {Miranda}, R., \& {Lai}, D. 2019, ApJ, 871, 84

\bibitem[{{Olofsson} {et~al.}(2019){Olofsson}, {Khouri}, {Maercker}, {Bergman}, {Doan}, {Tafoya}, {Vlemmings}, {Humphreys}, {Lindqvist}, {Nyman}, \& {Ramstedt}}]{HD101584}
{Olofsson}, H., {Khouri}, T., {Maercker}, M., {et~al.} 2019, A\&A, 623, A153

\bibitem[{{Oomen} {et~al.}(2020){Oomen}, {Pols}, {Van Winckel}, \& {Nelemans}}]{Oomen_2020}
{Oomen}, G., {Pols}, O., {Van Winckel}, H., \& {Nelemans}, G. 2020, A\&A, 642, A234

\bibitem[{{Oomen} {et~al.}(2019){Oomen}, {Van Winckel}, {Pols}, \& {Nelemans}}]{Oomen_2019}
{Oomen}, G., {Van Winckel}, H., {Pols}, O., \& {Nelemans}, G. 2019, A\&A, 629, A49

\bibitem[{{Oomen} {et~al.}(2018){Oomen}, {Van Winckel}, {Pols}, {Nelemans}, {Escorza}, {Manick}, {Kamath}, \& {Waelkens}}]{Oomen_2018}
{Oomen}, G., {Van Winckel}, H., {Pols}, O., {et~al.} 2018, A\&A, 620, A85

\bibitem[{{Paxton} {et~al.}(2011){Paxton}, {Bildsten}, {Dotter}, {Herwig}, {Lesaffre}, \& {Timmes}}]{MESA_2011}
{Paxton}, B., {Bildsten}, L., {Dotter}, A., {et~al.} 2011, ApJS, 192, 3

\bibitem[{{Paxton} {et~al.}(2013){Paxton}, {Cantiello}, {Arras}, {Bildsten}, {Brown}, {Dotter}, {Mankovich}, {Montgomery}, {Stello}, {Timmes}, \& {Townsend}}]{MESA_2013}
{Paxton}, B., {Cantiello}, M., {Arras}, P., {et~al.} 2013, ApJS, 208, 4

\bibitem[{{Paxton} {et~al.}(2015){Paxton}, {Marchant}, {Schwab}, {Bauer}, {Bildsten}, {Cantiello}, {Dessart}, {Farmer}, {Hu}, {Langer}, {Townsend}, {Townsley}, \& {Timmes}}]{MESA_2015}
{Paxton}, B., {Marchant}, P., {Schwab}, J., {et~al.} 2015, ApJS, 220, 15

\bibitem[{{Paxton} {et~al.}(2018){Paxton}, {Schwab}, {Bauer}, {Bildsten}, {Blinnikov}, {Duffell}, {Farmer}, {Goldberg}, {Marchant}, {Sorokina}, {Thoul}, {Townsend}, \& {Timmes}}]{MESA_2018}
{Paxton}, B., {Schwab}, J., {Bauer}, E.~B., {et~al.} 2018, ApJS, 234, 34

\bibitem[{{Paxton} {et~al.}(2019){Paxton}, {Smolec}, {Schwab}, {Gautschy}, {Bildsten}, {Cantiello}, {Dotter}, {Farmer}, {Goldberg}, {Jermyn}, {et~al.}}]{MESA_2019}
{Paxton}, B., {Smolec}, R., {Schwab}, J., {et~al.} 2019, ApJS, 243, 10

\bibitem[{{Rafikov}(2016)}]{Rafikov_2016}
{Rafikov}, R.~R. 2016, AJ, 830, 8

\bibitem[{{Rao} \& {Giridhar}(2014)}]{Rao_2014}
{Rao}, S.~S., \& {Giridhar}, S. 2014, \rmxaa, 50, 49

\bibitem[{{Reimers}(1975)}]{Reimers}
{Reimers}, D. 1975, MSRSL, 8, 369

\bibitem[{{Sabin} {et~al.}(2015){Sabin}, {Wade}, \& {Lebre}}]{Sabin_2015}
{Sabin}, L., {Wade}, G., \& {Lebre}, A. 2015, MNRAS, 446, 1988

\bibitem[{{Sch{\"o}nberner}(1979)}]{Schonberner_1979}
{Sch{\"o}nberner}, D. 1979, A\&A, 79, 108

\bibitem[{{Sch{\"o}nberner}(1987)}]{Schonberner_1987}
{Sch{\"o}nberner}, D. 1987, in The Second Conference on Faint Blue Stars: International Astronomical Union, Colloquium no. 95, 201--210

\bibitem[{{Schr{\"o}der} \& {Cuntz}(2007)}]{Updated_Reimers}
{Schr{\"o}der}, K.~P., \& {Cuntz}, M. 2007, A\&A, 465, 593

\bibitem[{{Tosi} {et~al.}(2022){Tosi}, {Dell\'{}Agli, F.}, {Kamath, D.}, {Ventura, P.}, {Van Winckel, H.}, \& {Marini, E.}}]{Tosi_et_al_2022}
{Tosi}, S., {Dell\'{}Agli, F.}, {Kamath, D.}, {et~al.} 2022, A\&A, 668, A22

\bibitem[{{Van Winckel}(1997)}]{Van_Winckel_1997}
{Van Winckel}, H. 1997, A\&A, 319, 561

\bibitem[{Van~Winckel(2003)}]{VanWinckel_2003}
Van~Winckel, H. 2003, ARA\&A, 41, 391

\bibitem[{{Van Winckel}(2014)}]{van_winckel_2014}
{Van Winckel}, H. 2014, Proc. Int. Astron. Union, 9, 180–186

\bibitem[{Van~Winckel(2018)}]{Van_Winckel_2018}
Van~Winckel, H. 2018, Proc. Int. Astron. Union, 92–105

\bibitem[{{Van Winckel}(2019)}]{VanWinckel_2019_review}
{Van Winckel}, H. 2019, in The Impact of Binary Stars on Stellar Evolution (Cambridge University Press), 92–105

\bibitem[{Van~Winckel {et~al.}(2012)Van~Winckel, Hrivnak, Gorlova, Gielen, \& Lu}]{van_winckel_2012}
Van~Winckel, H., Hrivnak, B.~J., Gorlova, N., Gielen, C., \& Lu, W. 2012, A\&A, 542, A53

\bibitem[{{Vassiliadis} \& {Wood}(1994)}]{stellar_wind}
{Vassiliadis}, E., \& {Wood}, P.~R. 1994, ApJS, 92, 125

\bibitem[{{Ventura} {et~al.}(2008){Ventura}, {D'Antona}, \& {Mazzitelli}}]{Ventura_2008}
{Ventura}, P., {D'Antona}, F., \& {Mazzitelli}, I. 2008, \apss, 316, 93

\bibitem[{{Ventura} {et~al.}(2020){Ventura}, F., {Lugaro}, {Romano}, {Tailo}, \& A.}]{Ventura_2020}
{Ventura}, P., F., D., {Lugaro}, M., {et~al.} 2020, A\&A, 641, A103

\bibitem[{{Ventura} {et~al.}(2018){Ventura}, {Karakas}, {Dell'Agli}, {García–Hernández}, \& {Guzman-Ramirez}}]{Ventura_2018}
{Ventura}, P., {Karakas}, A., {Dell'Agli}, F., {García–Hernández}, D.~A., \& {Guzman-Ramirez}, L. 2018, MNRAS, 475, 2282

\end{thebibliography}

%%%%% APPENDIX %%%%%
\appendix
\section{Spectral energy distributions}
\label{Appendix_SEDs}

\begin{figure*}
    \captionsetup[subfigure]{aboveskip=-1pt,belowskip=-1pt, oneside,margin={0.5cm,0cm}}
    \centering
    \begin{subfigure}{\textwidth}
    \centering
        \begin{subfigure}{0.47\textwidth}
         \caption{EP Lyr}
            \raisebox{0.02\height}{\includegraphics[width=\textwidth]{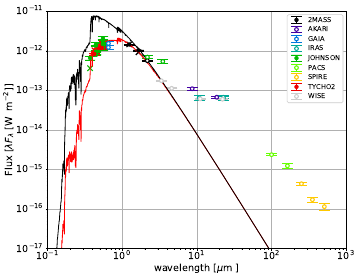}}
        \end{subfigure}
        %\hfill
        \hspace{5mm}
        \begin{subfigure}{0.475\textwidth}
            \caption{HP Lyr}
            \vspace{3pt}
            \includegraphics[width=\textwidth]{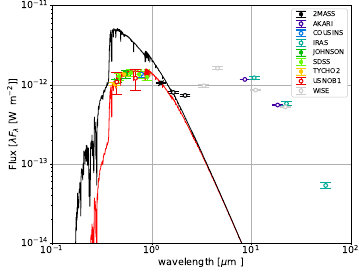}
        \end{subfigure}
    \end{subfigure}
    
    \vspace{4mm} % Adjust vertical space
    
    \begin{subfigure}{\textwidth}
    \centering
        \begin{subfigure}{0.47\textwidth}
            \caption{IRAS~17038-4815}
            \includegraphics[width=\textwidth]{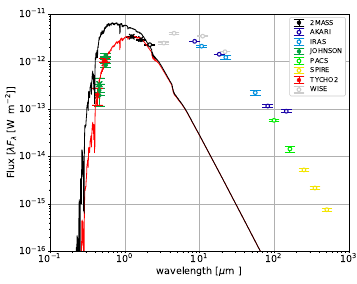}
        \end{subfigure}
        %\hfill
        \hspace{5mm}
        \begin{subfigure}{0.48\textwidth}
            \caption{IRAS~09144-4933}
            \vspace{3pt}
            \includegraphics[width=\textwidth]{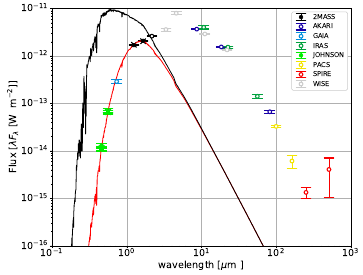}
        \end{subfigure}
    \end{subfigure}
    
    \vspace{4mm} % Adjust vertical space
    
    \begin{subfigure}{\textwidth}
    \centering
        \begin{subfigure}{0.47\textwidth}
            \caption{HD~131356}
            \vspace{2pt}
            \includegraphics[width=\textwidth]{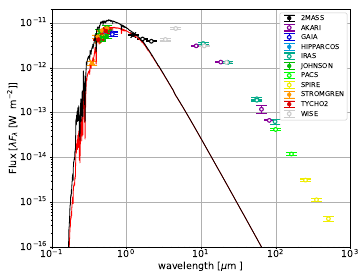}
        \end{subfigure}
        %\hfill
        \hspace{5mm}
        \begin{subfigure}{0.475\textwidth}
            \caption{SX Cen}
            \vspace{-1pt}
            \includegraphics[width=\textwidth]{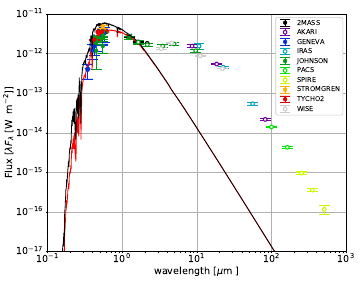}
        \end{subfigure}
    \end{subfigure}
    
    \caption{Spectral energy distributions (SEDs) from \protect\cite{Oomen_2018} for the chemically-depleted post-AGB binary stars used in this work. The observed spectrum (red curve) and dereddened photospheric model (black curve, scaled to the object) are shown. Photospheric observations were obtained at different pulsation phases by various surveys (indicated by different coloured symbols); for expansion on acronyms used different photometric surveys, we refer the reader to the relevant instrument papers. All objects (except EP~Lyr) present a distinct near-infrared (near-IR) excess, indicative of hot dust in a stable circumbinary disk. The lack of IR excess in the SED of EP~Lyr is characteristic to a transition-type disk, which have large dust-free inner-disk cavities.}
    \label{SED_plots}
\end{figure*}

In Figure~\ref{SED_plots}, we provide the spectral energy distribution (SED) plots for our post-AGB binary star sample (see Section~\ref{postAGB_sample}). The SEDs of post-AGB binary stars serve as the earliest and strongest source of direct observational evidence for a dusty circumbinary disk. The SEDs presented in Figure~\ref{SED_plots} were obtained from \cite{Oomen_2018}, and were prepared using photometry from various surveys, spanning blue, optical, and near-IR wavelengths, to fit the observed photosphere of the post-AGB primary star. For further details on the SED fitting procedure, we refer the reader directly to \cite{Oomen_2018}.

%\FloatBarrier

\section{Revised stellar and binary properties for EP~Lyr and HP~Lyr using Gaia DR3}
\label{Appendix_GDR3}

This Appendix demonstrates how the stellar and binary properties of post-AGB systems change with ongoing updates on distance measurements. In Table~\ref{update_GDR3_appendix}, we provide \textit{updated} observed and derived parameters for post-AGB binary stars EP~Lyr and HP~Lyr (i.e., RV Tauri pulsators), based on \textit{Gaia} DR3 distances from \cite{GaiaDR3_dist_BailerJones2021}. These stars exhibit significant variability, which further complicates precise measurements of key stellar parameters, making them particularly useful for illustrating the effects of revised distance estimates on their derived properties. Effective temperatures in Table~\ref{update_GDR3_appendix} were obtained from \cite{Maksym_2023b} for EP~Lyr and \cite{Oomen_2019} for HP~Lyr, while photospheric luminosities were determined by \cite{Meghna_2024} through the PLC-relation. The current (post-AGB) mass of each star was estimated using the core mass-luminosity relation of \cite{stellar_wind}, and the current radius was calculated from the derived luminosity and observed effective temperature. The upper-limit Roche-lobe radii at periastron (in Table~\ref{update_GDR3_appendix}) were computed using the standard \cite{eggleton_1983} formula, assuming derived companion masses from \cite{Oomen_2018,Oomen_2020}.

\begin{table}[hbt!]
    \centering
    \caption{Updated observed and derived properties for EP~Lyr and HP~Lyr, based on \textit{Gaia} DR3 distances. Effective temperatures ($T_{\rm{eff}}$) are taken from \protect\cite{Maksym_2023b} for EP~Lyr and \protect\cite{Oomen_2019} for HP~Lyr, while photospheric luminosities, derived via the PLC-relation, are from \protect\cite{Meghna_2024}. Further provided are the estimated current (post-AGB) mass and stellar radius, along with values for orbital period, eccentricity, and projected semi-major axis ($a_{\rm{b}}\sin i$) from \protect\cite{Oomen_2018,Oomen_2020}. The upper-limit Roche-lobe radius at periastron ($R_{\rm{L,,peri}}$) is additionally given.}
    \setlength{\tabcolsep}{4pt}
    \renewcommand{\arraystretch}{1.0}
    \resizebox{0.85\textwidth}{!}{\begin{tabular}{p{23mm} >{\centering\arraybackslash}p{20mm} >{\centering\arraybackslash}p{20mm}}
         \hline
         \hline 
         \rule{0pt}{2mm}Star Name & EP Lyr & HP Lyr \\[0.2mm]
         \hline 
         $T_{\rm{eff}}$ (K)  & $6270\pm160$ & $6300\pm250$ \\
         %Distance (kpc)  & $4.1\,^{+0.4}_{-0.5}$ & $10.7\,^{+1.3}_{-1.4}$ \\[0.1mm]
         Luminosity ($10^3$~\Lsun)  & $5.7\,^{+3.0}_{-4.5}$ & $9.5\,^{+6.5}_{-5.3}$ \\
         \referee{Current mass (\Msun)}  & \referee{$0.60$} & \referee{$0.67$} \\
         Radius (\Rsun)  & $52\pm4$ & $85\pm7$ \\
         Period (days)   & $1151\pm14$   & $1818\pm80$ \\
         Eccentricity    & $0.39\pm0.09$ & $0.20\pm0.04$ \\
         $a_{\rm{b}}\sin i$ (AU) & $1.30\pm0.12$ & $1.27\pm0.06$ \\  
         $R_{\rm{L,\,peri}}$ (\Rsun)  & $75\pm15$ & $76\pm11$  \\
         \hline
    \end{tabular}}
    \label{update_GDR3_appendix}
\end{table}
%Updated observed and derived properties, based on \textit{Gaia} DR3 distances \protect\cite[from][]{GaiaDR3_dist_BailerJones2021}, for post-AGB binary stars EP~Lyr and HP~Lyr. The post-AGB star's effective temperature ($T_{\rm{eff}}$), PLC-derived photospheric luminosity \protect\cite[from][]{Meghna_2024}, estimated current mass, and approximate stellar radius, are provided. Orbital periods, eccentricities, and projected semi-major axes ($a_{\rm{b}}\sin i$), are from \protect\cite{Oomen_2018,Oomen_2020}. The estimated Roche-lobe radii at periastron ($R_{\rm{L,\,peri}}$) is additionally given.

\section{Impact of late thermal pulses on evolving post-AGB stars}
\label{Appendix_LTP}

\referee{In Table~\ref{0pt55Mo_extfactors_table}, we present the post-AGB evolution parameters for the 0.55~\Msun\ models, which were found to experience a late thermal pulse \cite[LTP;][]{Schonberner_1979, Herwig_2001}. During an LTP, renewed helium burning alters the star's radius and effective temperature, significantly impacting the post-AGB lifetime \cite[]{Blocker_2001}. The timing of an LTP depends on the moment during the thermal pulse cycle at which the star leaves the AGB, with this departure being driven in our models by the rapid removal of the stellar envelope to simulate a strong binary interaction (see Section~\ref{postAGBmodelling_section}). LTPs occur in around 20\% of post-AGB stars, and whether a stellar model experiences an LTP is stochastic \cite[though LTPs can be artificially induced by modifying the timing of envelope removal;][]{Iben_et_al_1983}.}

\begin{table}[hbt!]
    \centering
    \caption{Post-AGB evolution timescales, expressed in years, and extension factors (enclosed by square brackets) for the accreting 0.55~\Msun\ models, given different initial accretion rates, $\dot{M}(0)$, and disk masses, $M_{\rm{d}}$. Values are computed similarly to those presented in Section~\ref{effects_on_properties} (Table~\ref{PAGB_ext_factors_results_table}) for the 0.60~\Msun\ models, though come with a caveat \textendash\ see text. The post-AGB evolution timescale ($\mathlarger{\mathlarger{\tau}}_{\rm{pAGB}}$) corresponds to the model evolution time between $T_{\rm{eff}}=$~4000~K and $T_{\rm{eff}}=$~25~000~K. The post-AGB phase extension factor is computed as $\mathlarger{\mathlarger{\tau}}_{\rm{pAGB}}/\mathlarger{\mathlarger{\tau}}_{\rm{\scriptsize{norm}}}$, where $\mathlarger{\mathlarger{\tau}}_{\rm{\scriptsize{norm}}}$ is the post-AGB evolution timescale of the non-accreting model (noted below the table).}
    \setlength{\tabcolsep}{1.5pt}
    \renewcommand{\arraystretch}{1.05}
    \resizebox{\linewidth}{!}{\begin{tabular}{p{23mm} >{\centering\arraybackslash}p{15mm} >{\centering\arraybackslash}p{15mm} >{\centering\arraybackslash}p{15mm} >{\centering\arraybackslash}p{15mm}}
         \hline
         \hline
         \rule{0pt}{2mm}$\dot{M}(0)$ (\,\Mdot\,) & $1\times10^{-8}$ & $1\times10^{-7}$ & $5\times10^{-7}$ & $1\times10^{-6}$ \\[0.5ex]
         \hline
         %\\[-4ex]
         $M_{\rm{d}}\,=\,3\times10^{-2}$~\Msun   &  $3632\, \left[1.00\right]$ & $4672 \, \left[1.29\right]$ & $7850 \, \left[2.16\right]$ & $7851 \, \left[2.16\right]$ \\
         $M_{\rm{d}}\,=\,1\times10^{-2}$~\Msun   &  $3632 \, \left[1.00\right]$ & $4509 \, \left[1.24\right]$  & $7848 \, \left[2.16\right]$  & $7848 \, \left[2.16\right]$  \\
         $M_{\rm{d}}\,=\,7\times10^{-3}$~\Msun   &  $3632 \, \left[1.00\right]$ & $4429 \, \left[1.22\right]$  & $7843 \, \left[2.16\right]$  & $7844 \, \left[2.16\right]$  \\[1ex]
         \hline
         \multicolumn{5}{p{\linewidth}}{\textbf{Note.} The non-accreting $0.55$~\Msun\ model has $\tau_{\rm{\scriptsize{norm}}}=3632$~years.} \\
    \end{tabular}}
    \label{0pt55Mo_extfactors_table}
\end{table}

\begin{figure}[hbt!]
    \centering
    \includegraphics[width=0.95\linewidth]{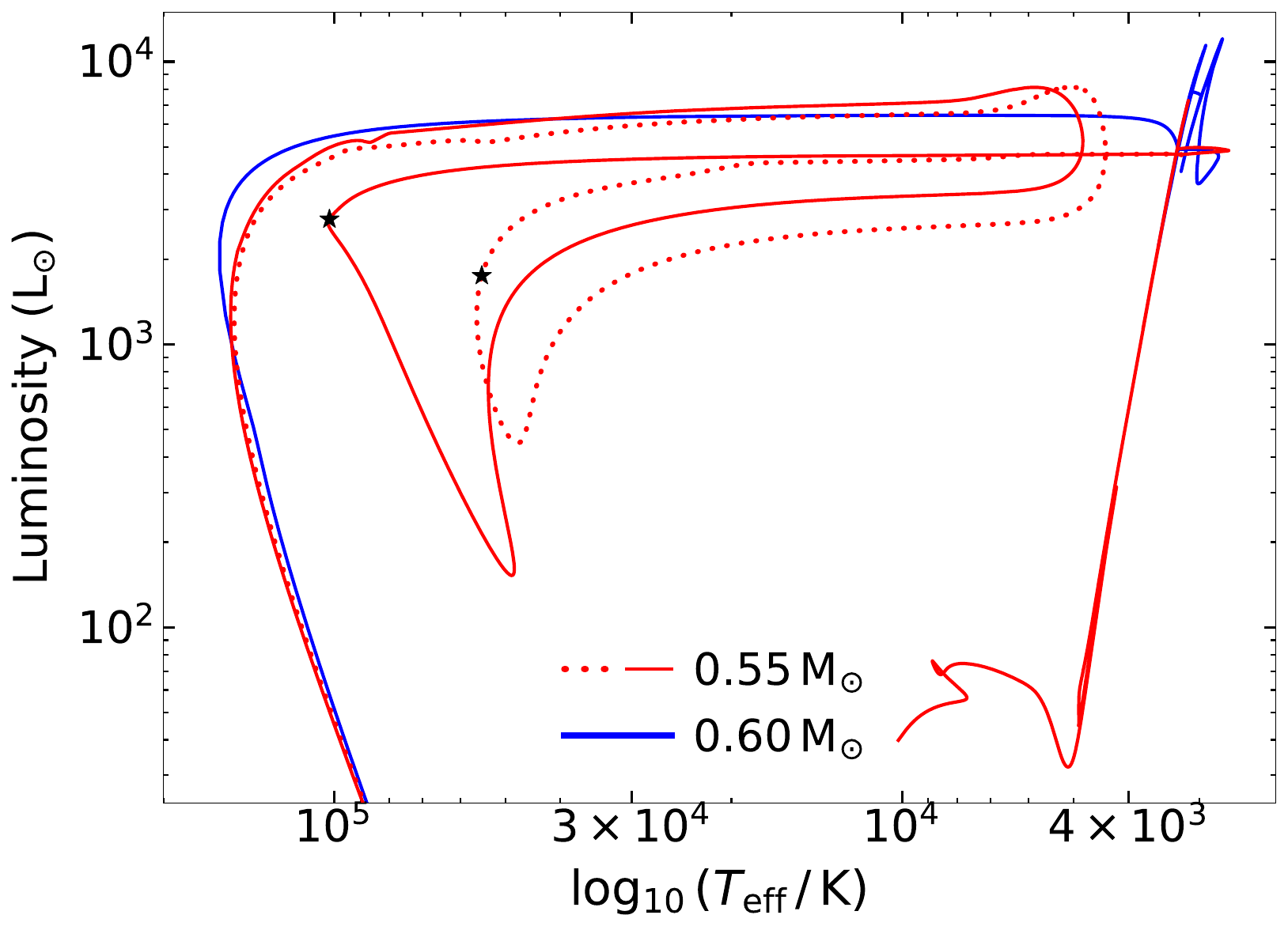}
    \caption{Hertzsprung-Russel diagram showing the \mesa\ evolutionary tracks of the non-accreting 0.55~\Msun\ (red solid) and 0.60~\Msun\ (blue solid) post-AGB models, and the most rapidly accreting ($\dot{M}(0)=10^{-6}$~\Mdot) 0.55~\Msun\ post-AGB model with disk mass $3\times10^{-2}$~\Msun\ (red dotted). The stellar models presented were evolved from the main sequence, with an initial mass of 2.5~\Msun, and initial solar metallicity. A late thermal pulse (LTP) is shown to occur in both the non-accreting and accreting 0.55~\Msun\ post-AGB models, which temporarily brings the star back to the post-AGB phase. A black star marks the peak of the LTP in both the non-accreting and accreting 0.55~\Msun\ models, around 8000~years into the post-AGB phase.}
    \label{HRD_appplot}
\end{figure}

When an LTP occurs, a star's temperature evolution and consequently, its post-AGB timescale, differ substantially from those of a star that did not experience an LTP, making the direct comparison of stellar models with and without an LTP inappropriate. Therefore, the post-AGB evolution timescales for the 0.55~\Msun\ models in Table~\ref{0pt55Mo_extfactors_table}, which are influenced by an LTP, cannot be compared to those of the 0.60~\Msun\ models in Section~\ref{effects_on_properties} (Table~\ref{PAGB_ext_factors_results_table}), where not LTP occurs. In contrast, both the non-accreting and accreting 0.60~\Msun\ models do not experience an LTP, making them directly comparable (see Section~\ref{evol_effects_section}). The occurrence of an LTP in the 0.55~\Msun\ models, while not in the 0.60~\Msun\ models, is shown clearly in Figure~\ref{HRD_appplot}.

\begin{figure}[hbt!]
    \centering
    \includegraphics[width=0.96\linewidth]{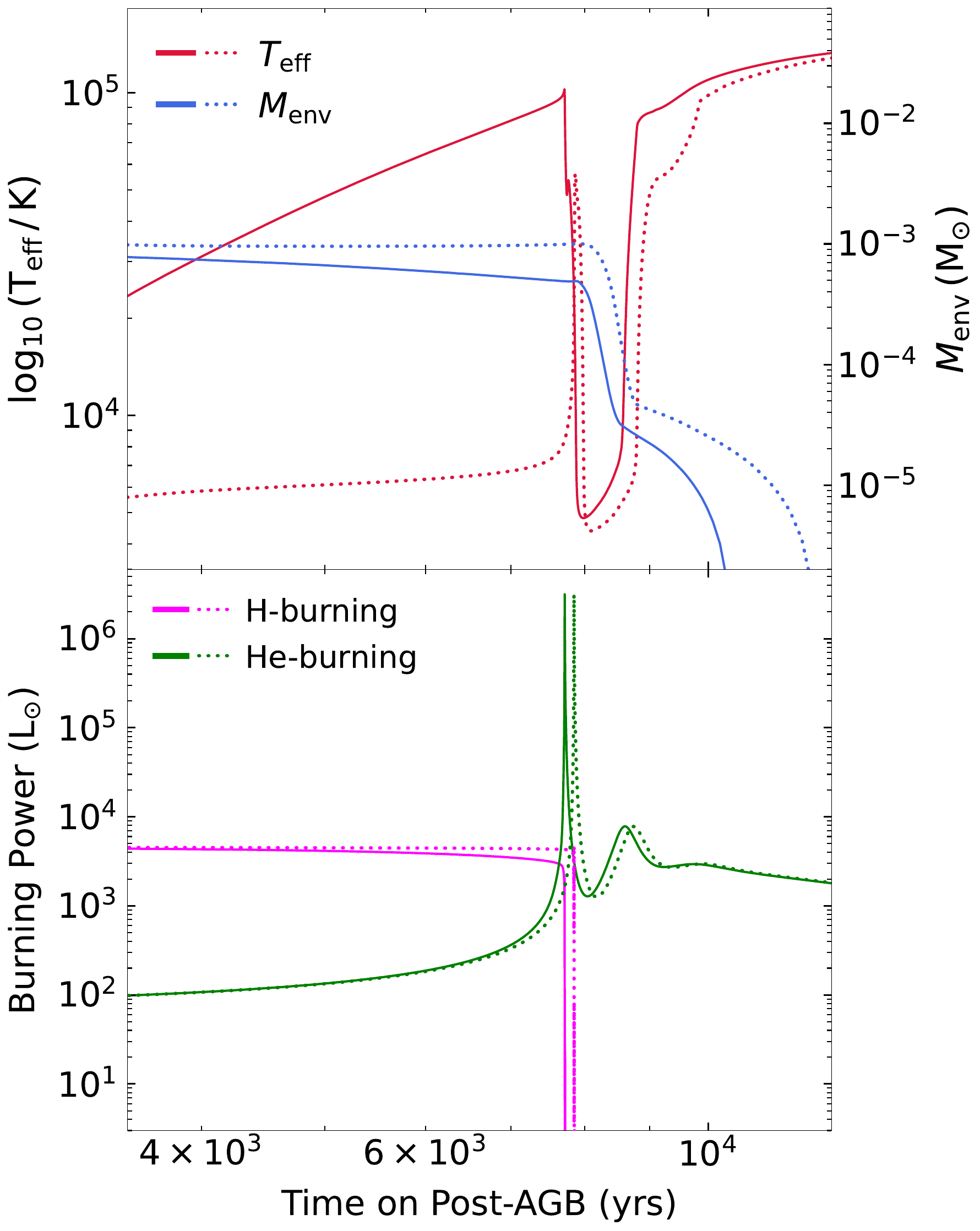}
    \caption{Evolution of effective temperature ($T_{\rm{eff}}$), envelope mass ($M_{\rm{env}}$), and nuclear burning power (from hydrogen and helium) during the late thermal pulse in the 0.55~\Msun\ models. Solid lines represent the non-accreting 0.55~\Msun\ model, while dotted lines correspond to the most rapidly accreting ($\dot{M}(0)=10^{-6}$~\Mdot) 0.55~\Msun\ model with disk mass $3\times10^{-2}$~\Msun.}
    \label{LTP_params_appplot}
\end{figure}

Figure~\ref{LTP_params_appplot} shows the evolution of effective temperature, envelope mass, and nuclear burning rate (due to hydrogen and helium) during the LTP in the non-accreting 0.55~\Msun\ model, and the most rapidly accreting ($\dot{M}(0)=10^{-6}$~\Mdot) 0.55~\Msun\ model with a disk mass of $3\times10^{-2}$~\Msun. The peak of the LTP, when helium-burning luminosity is highest, occurs around 8000~years into the post-AGB phase: at an effective temperature of $\sim10^{5}$~K in the non-accreting model and $\sim$~50~000~K in the accreting model, as indicated by a black star on the evolutionary tracks in Figure~\ref{HRD_appplot}. 

Both models in Figure~\ref{LTP_params_appplot} reach an effective temperature of 25~000~K (i.e., the end of the post-AGB phase) before the LTP reaches its peak at around 8000~years. This suggests that despite the occurrence of the LTP, the two models could still be compared. However, it is clear that while the accreting model retains a lower effective temperature for longer (due to accretion), it also exhibits a reduced luminosity (see Figures~\ref{HRD_appplot}~and~\ref{LTP_params_appplot}), indicating that the LTP influences the temperature evolution at earlier times than its peak at $\sim$~8000~years. Therefore, while the 0.55~\Msun\ models still provide useful insights for post-AGB evolution, their computed post-AGB timescales (and corresponding post-AGB phase extension factors) should be treated with caution, as the LTP significantly affects the temperature evolution upon which these timescales depend.

The interplay between accretion and the occurrence of an LTP is certainly interesting. An investigation of whether LTPs become more or less prevalent as a result of accretion would be important to determine their impact on the emergence and characteristics of the post-AGB binary class as a whole.

\section{Mixing depth of accreted gas in post-AGB model envelopes}
\label{Appendix_Menv}

\begin{figure}[hbt!]
    \centering
    \includegraphics[width=0.95\linewidth]{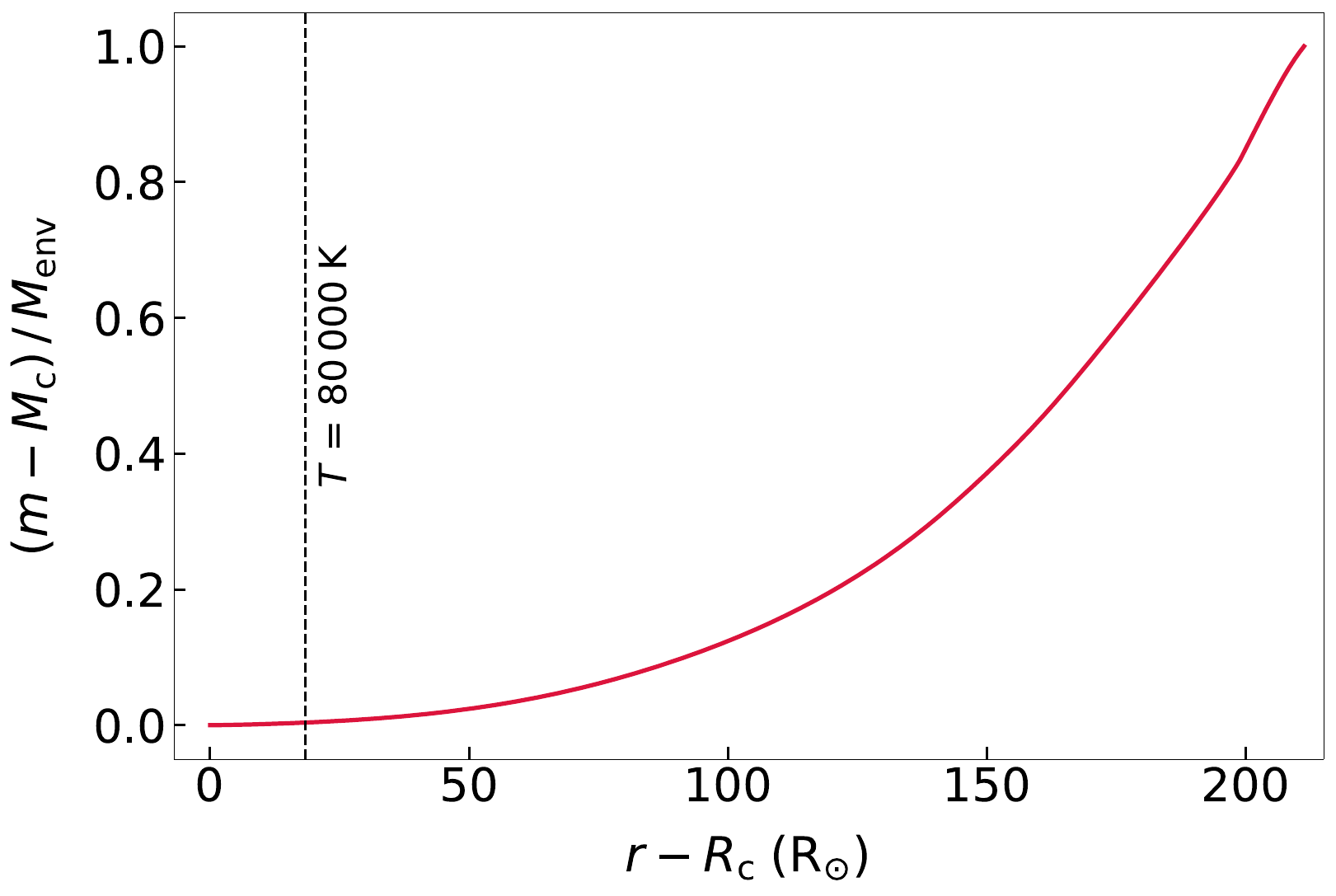}
    \caption{Envelope mass fraction as a function of radial coordinate, from the base of the convective envelope to the surface of the star, showing the depth to which accreted material was mixed (dashed line). Envelope mass ($M_{\rm{env}}$) was computed by subtracting the core mass ($M_{\rm{c}}$, initially $\sim$~0.53~\Msun) from the total stellar mass at each mass coordinate $m$, and has been further normalised to the mass at the convective envelope base. The radial coordinate was computed as the difference between the radial coordinate for the star, $r$, and the helium core radius ($R_{\rm{c}}$, initially $\sim$~0.03~\Rsun).}
    \label{env_MoverR_mixplot}
\end{figure}

Figure~\ref{env_MoverR_mixplot} illustrates the depth to which accreted material is mixed into the convective envelope of the post-AGB models used in this study. Specifically, accreted gas from a circumbinary disk was mixed down to regions in the convective envelope where $T\lesssim$~80~000~K (see Section~\ref{depl_method}); as shown in Figure~\ref{env_MoverR_mixplot}, this reaches near the base of the envelope, mixing through approximately 100\% of the envelope mass (i.e., $\sim$~0.02~\Msun\ at the start of the post-AGB phase). As noted by \cite{Oomen_2019}, the exact choice of mixing depth for accreted material has minimal impact on the resulting depletion of post-AGB stars at effective temperatures below $\sim$~7000~K, such as those used in this study.

\end{document}